\documentclass	[english,aps,prb,reprint, superscriptaddress]{revtex4-2}
\usepackage[T1]{fontenc}
\usepackage[latin9]{inputenc}
\usepackage{geometry}
\usepackage{verbatim}
\usepackage{float}
\usepackage{mathrsfs}
\usepackage{amsmath}
\usepackage{amssymb}
\usepackage{graphicx}
\usepackage{esint}
\usepackage{multirow}
\usepackage{xcolor}
\usepackage[normalem]{ulem}
\makeatletter
\usepackage{grffile}
\usepackage{amsmath}
\usepackage{tikz}

\renewcommand{\fnum@figure}{FIG. \thefigure}

\makeatother

\usepackage{babel}

\begin{document}

\author{Yonatan Messica}
\affiliation{
	Department of Physics, Bar-Ilan University, Ramat Gan, 52900, Israel
}
\author{Pavel M. Ostrovsky}
\affiliation{
	Max Planck Institute for Solid State Research, Heisenbergstrasse 1, 70569, Stuttgart, Germany
}
\affiliation{
	L. D. Landau Institute for Theoretical Physics RAS, 142432 Chernogolovka, Russia
}
\author{Dmitri B. Gutman}
\affiliation{
	Department of Physics, Bar-Ilan University, Ramat Gan, 52900, Israel
}

\begin{abstract}
We study heat transport in a Weyl semimetal with broken time-reversal symmetry in the hydrodynamic
regime.
At the neutrality point, the longitudinal heat conductivity is governed by the momentum relaxation (elastic) time, while the longitudinal electric conductivity is controlled by the inelastic scattering time. In the  hydrodynamic regime this  leads to a large longitudinal Lorenz ratio. As the chemical potential is tuned away from the neutrality point, the longitudinal Lorenz ratio decreases because of suppression of the heat conductivity by the Seebeck effect.
The Seebeck effect (thermopower) and the open circuit heat conductivity are intertwined with the electric conductivity. The magnitude of the Seebeck tensor is parametrically enhanced, compared to the non-interacting model, in a wide parameter range. While the longitudinal component of Seebeck response decreases with increasing  electric anomalous Hall conductivity $\sigma_{xy}$, the transverse  component depends on $\sigma_{xy}$ in a non-monotonous way. Via its effect on the Seebeck response, large $\sigma_{xy}$  enhances the longitudinal Lorenz ratio at a finite chemical potential.
At the neutrality point, the transverse heat conductivity is determined by the Wiedemann-Franz law.
Increasing the distance from the neutrality point, the transverse heat conductivity is enhanced by the transverse Seebeck effect and follows its non-monotonous dependence on $\sigma_{xy}$.
\end{abstract}

\title{Heat transport in Weyl semimetals in the hydrodynamic
regime}

\maketitle

\section{Introduction}

When electron-electron collisions are the fastest scattering mechanism,
a metal enters the hydrodynamic regime. In this regime, electrons
reach a local thermal equilibrium and flow in a collective manner,
described by slowly varying degrees of freedom. This gives rise to
a variety of effects known in the context of classical fluids. 
Electron flow in this  regime is  controlled by   viscosity and  characterized by the emergence 
of  large scale patterns \cite{Levitov2016}.
 While the idea of hydrodynamic electrons
was conceptualized decades ago \cite{Gurzhi1963}, advances in fabrication
technology have enabled experimental realization in a variety of systems
in recent years \cite{Ghahari2016,Moll2016,Crossno2016,Gooth2018,Jaoui2018,Vool2021,Jaoui2021,Amit2022},
sparking a large interest in the field.

Thermal transport is another phenomenon where the difference between non-interacting metals and metals in the hydrodynamic regime is dramatic.
In a non-interacting metal, the thermal
conductivity and electric conductivity are related via the Wiedemann-Franz (WF) law, stating that their ratio divided by the temperature, the Lorenz ratio $\mathcal{L}_{\alpha\beta} \equiv T^{-1}\kappa_{\alpha \beta}/ \sigma_{\alpha \beta}$, is given by $\mathcal{L}_{0}=\pi^{2}/\left(3e^{2}\right)$.
An analogous relation between the electric conductivity and thermopower (Seebeck coefficient) is known as the Mott relation \cite{ashcroft1976solid}.
In the presence of interactions, inelastic collisions between the electrons lead to a separation of the electric and thermal
degrees of freedom, breaking the relation between thermal and electric conductivity.

In graphene exhibiting hydrodynamic transport, for example, the Lorenz ratio greatly exceeds the WF result at the charge
neutrality point, while going below it at higher carrier densities
\cite{Crossno2016}. Similarly, the  thermopower in graphene is also enhanced \cite{Ghahari2016},
exceeding the value predicted by the Mott relation. Besides fundamental interest,  
enhancement of thermopower may be favorable for applications
such as thermoelectric energy harvesting \cite{Fu2020,Andersen2019}.

There are several ways to describe thermal transport. 
To define thermal conductivity in a meaningful way, it is important to
specify the experimental setup in which thermal currents are measured.
The standard choices  correspond to either  
a zero electric current  (open circuit boundary condition) or a zero electric field.
Due to the cross electric-thermal responses, these two setups lead
to different thermal conductivities. While for non-interacting metals
the difference is typically negligible due to the smallness of the
Seebeck effect \cite{ashcroft1976solid}, in the hydrodynamic regime
they can be drastically different \cite{Lucas2018}. As we will describe
in this work, a combination of a strong Seebeck effect in the hydrodynamic
regime together with transverse anomalous Hall transport will have important consequences
on the open circuit heat conductivity.

In this work we will focus on the thermal transport in 
Weyl semimetals (WSMs). WSMs are 3D materials with topologically protected
band-crossing points known as Weyl nodes \cite{Yan2017, Armitage2018}. In the vicinity
of the Weyl nodes, the density of states is vanishing and the dispersion
is linear. The Weyl nodes are sources of Berry curvature, and hence
cannot be gapped out without two pairing nodes merging into one Dirac
node. Due to the Berry curvature, electrons acquire an anomalous velocity
perpendicular to an applied electric field \cite{Xiao2010}. In time-reversal symmetry
(TRS) breaking WSMs, the total Berry curvature of filled bands does
not vanish, giving rise to the anomalous Hall effect (AHE), which
is proportional to the distance between pairing Weyl nodes \cite{Burkov2011b}.

The Berry curvature also gives rise to anomalous thermoelectric transport,
manifested in the anomalous thermal Hall and Nernst effects \cite{Qin2011, Haldane2004}.
In the anomalous thermal Hall effect, a temperature
gradient induces thermal current in the transverse direction, and
in the anomalous Nernst effect, a transverse electric field emerges due 
to a temperature gradient. Remarkably, even though these anomalous effects have a topological origin,
the thermoelectric transport coefficients in zero temperature
non-interacting WSMs obey the same relations as
in normal metals, namely the WF law and the Mott relation \cite{Xiao2006, Qin2011, Gorbar2017}.

The distinct properties of WSMs have implications for the behavior in the hydrodynamic regime.
In particular,  for scales where mixing between  different  Weyl nodes can be neglected, 
the system is  made of chiral  fluids.  Each fluid is made of the electronic states corresponding to a given node and inherits the chirality of the node.
 This gives rise to unique effects, such as collective anomalous Hall waves \cite{Gorbar2018}
and thermal magnetoresistance, manifesting the axial-gravitational
anomaly \cite{Lucas2016}. Importantly, the semimetallic nature of
WSMs may be beneficial for establishing the hydrodynamic regime, since
the absence of a Fermi surface makes the screening of Coulomb interactions
much weaker than in a normal metal \cite{Abrikosov1971}. Indeed,
one of the successful realizations of the hydrodynamic regime was
done for the WSM tungsten diphosphide (WP$_2$), where an exceptionally
low value of the Lorenz ratio was achieved \cite{Gooth2018}.

In this work, we investigate the thermoelectric transport in a TRS-breaking
WSM in the hydrodynamic regime, at the vicinity of the Weyl nodes.
We calculate the thermoelectric conductivities using the Boltzmann equation formalism. 
We find that electric and heat conductivity are differently affected by inelastic scattering.
While the  anomalous electric Hall conductivity  in the hydrodynamic regime is the same as for non-interacting electrons,  the longitudinal and transverse heat conductivities are parametrically different in these two regimes.
 
We stress, that  these results are derived for the standard definition of the heat conductivity, i.e. for  the open circuit setup. In this case, imposing a temperature gradient gives rise to an electric field required to maintain a zero electric current.
The magnitude of the electric field response is controlled by the electric resistivity tensor.
Therefore, the value of the anomalous
electric Hall conductivity affects both the longitudinal and transverse heat
conductivities. Additionally, we find that the Seebeck coefficients (both longitudinal and transversal) are enhanced in the hydrodynamic regime, compared to the non-interacting one, and are 
nearly equal to the entropy per electric charge in a wide range of parameters.

It is worth mentioning that due to the Dirac-like spectrum, the electronic hydrodynamics in WSMs possess Lorentz invariance (rather than Galilean), and belong to the class of relativistic fluids \cite{Landau2003}. As such, they share a lot with other materials in this family, particularly with graphene \cite{Narozhny2017, Lucas2018, Narozhny2019}.
One may ask, what is the difference between a TRS-breaking WSM and graphene in an external magnetic field? As will be clear from this work,
the key distinction is due the origin of the transverse currents.

In a graphene-like system, the effects of an external magnetic field are captured by adding the Lorentz force to the hydrodynamic equations. It gives rise to cyclotron motion of the fluid, 
and finite Hall thermoelectric conductivities \cite{Hartnoll2007, Muller2008, Miiller2009}.
This is to be contrasted with transverse transport due to Berry curvature in WSMs. 
The Berry curvature induces an anomalous channel of transverse conductivity, which is not accounted for by the boost velocity field.
The anomalous currents do not affect the longitudinal electric and thermoelectric conductivities, 
and modify the longitudinal heat conductivity and Seebeck coefficient only via the resistivity tensor.
This gives rise to a qualitatively different behavior of the thermoelectric responses in TRS-breaking WSMs in the hydrodynamic regime compared to the relativistic magnetohydrodynamics.




\section{Model, Boltzmann and hydrodynamic equations, response coefficients}

\subsection{Model}

We focus on a minimal model for TRS-breaking, inversion-symmetric Weyl semimetal, containing
two nodes at $\mathbf{k}=\pm \hat{z}\Delta_k/2 $.
The magnitude of the momentum separation 
between the Weyl nodes $\Delta_k$ 
determines the anomalous Hall conductivity. Focusing on the low-energy
part of the spectrum, the non-interacting part of the Hamiltonian near each Weyl node reads ($\hbar=k_{B}=1$):

\begin{equation}
H_{\eta}=\eta v_{F}\mathbf{\sigma}\cdot \left( \mathbf{k} - \eta \frac{\mathbf{\Delta}_k}{2} \right) ,\label{eq:Hamiltonian}
\end{equation}

where $\eta=\pm 1$ corresponds to the chirality of the node. In the vicinity of the nodes there are two bands $b=\pm$
with the spectrum $\epsilon_{\eta b \mathbf{k}}=b v_F \vert \mathbf{k} - \eta \frac{\mathbf{\Delta}_k}{2} \vert$.
We use the index $l=\left(\eta, b \right)$ as a short notation for the node and the band.
The Dirac cones described by the Hamiltonian (\ref{eq:Hamiltonian}) are without tilt. 
In a more general model, the Dirac cones may be tilted along a specific axis in momentum space.
This is modeled by adding the term $\eta \mathbf{u}_t \cdot \left( \mathbf{k} - \eta \frac{\mathbf{\Delta}_k}{2} \right)$\
to the 
Hamiltonian \cite{Armitage2018}. Here, $\mathbf{u}_t$ is a tilt vector with dimensions of velocity.
In our analysis we focus on the simplest, but  already interesting case of untilted Dirac cones.

The electrons scatter off each other by Coulomb interactions with
typical lifetime $\tau^{\textrm{e-e}}$, which is assumed to be the
shortest scattering time in the system. Additionally, we include disorder,
which is diagonal in node and pseudo-spin space and is described by a Gaussian correlator,

\begin{equation}
\left\langle V(\mathbf{r})V(\mathbf{r}')\right\rangle _{\textrm{disorder avg.}}=\gamma\delta(\mathbf{r}-\mathbf{r}').\label{eq:Gaussian disorder}
\end{equation}


Throughout this paper, we assume that the quasiparticle description
is valid.  We  also disregard many-body renormalization effects.

\subsection{Derivation of hydrodynamic equations}

First, we briefly describe the derivation of hydrodynamic equations
for the electron fluid. The derivation is similar to the one done
for graphene, which also exhibits Dirac spectrum but in two dimensions. 
For the recent reviews on graphene in the hydrodynamic regime see Refs. \cite{Lucas2018,Narozhny2019}.
The hydrodynamic equations can be derived from the Boltzmann equation,
which is given by

\begin{equation}
\frac{\partial f}{\partial t}+\dot{\mathbf{r}}\cdot\mathbf{\nabla}_{r}f+\dot{\mathbf{k}}\cdot\mathbf{\nabla}_{k}f=I_{\textrm{e-e}}\left[f\right]+I_{\textrm{e-imp}}\left[f\right],\label{eq:Boltzmann equation}
\end{equation}

where $I_{\textrm{e-e}},I_{\textrm{e-imp}}$ are the collision integrals
for electron-electron and electron-impurity scattering, correspondingly.
If the electron-electron scattering time is the shortest time scale in the system, 
the e-e collision integral projects  the electron distribution function
to a local near-equilibrium distribution
\begin{equation}
\label{near_equilibrium}
f_{l\mathbf{k}}=n_{F}(x_{l\mathbf{k}})+\delta f_{l\mathbf{k}},
\end{equation}
where,
\begin{equation}
x_{l\mathbf{k}}=\frac{\epsilon_{l\mathbf{k}}-\mu(\mathbf{r},t)-\mathbf{u}(\mathbf{r},t)\cdot\mathbf{k}}{T(\mathbf{r},t)}.
\end{equation}

The Fermi-Dirac part $n_{F}(x)\equiv\left(1+\exp(x)\right)^{-1}$
describes the zero-modes of the e-e collision integral, with the quantities
$\mu,\mathbf{u},T$ corresponding to the slowly changing modes of particle, momentum and
energy densities. The second term in Eq. (\ref{near_equilibrium}), 
$\delta f_{l\mathbf{k}}$, 
accounts for the dissipative part of the distribution function, corresponding
to the finite modes which are relaxed by the electron-electron scatterings.
We note that by assuming only three zero-modes, we are neglecting long-lived (but not strictly conserved) modes. One such mode is the imbalance mode,
corresponding to different chemical potentials for the electron and hole bands \cite{Foster2009, Lucas2018}.
This mode would give rise to finite-size corrections for the conductivities and is beyond the scope of this work.
Additionally, one may study the chiral imbalance mode, where the two Weyl nodes have different chemical potential.
This mode can be excited by having parallel magnetic and electric fields, producing chiral charge \cite{Son2013, Gorbar2018}.

The hydrodynamic equations describe the variation of the zero-modes
on the longer scale, perturbatively in the external forces and gradients.
The conservation equations for charge, momentum and energy densities ($x=n,\mathbf{\pi},n_{\epsilon}$)
are then obtained by multiplying
the Boltzmann equation with $y_{l\mathbf{k}}=1,\mathbf{k},\epsilon_{l\mathbf{k}}$,
integrating with respect to momentum and summing over the bands and nodes.

From the momentum and energy conservation equations, we obtain the
Euler equation for the boost velocity $\mathbf{u}$ (Appendix
\ref{sec:Euler equation derivation}),

\begin{widetext}
\begin{eqnarray}&&
\left(\frac{\partial}{\partial t}+\mathbf{u}\cdot\mathbf{\nabla}_{r}\right)\mathbf{u}+\frac{v_{F}^{2}}{w}\left[n\left(-e\mathbf{E}+\mathbf{\nabla}_{r}\mu\right)+s\mathbf{\nabla}_{r}T\right]+\frac{\mathbf{u}}{w}\left[n\left(e\mathbf{E}\cdot\mathbf{J}^{n}+\frac{\partial\mu}{\partial t}\right)+s\frac{\partial T}{\partial t}\right]
\nonumber \\&&
=\frac{1}{w}\sum_{l}\intop\left(d\mathbf{k}\right)\mathbf{k}I_{\mathrm {e-imp}}\left[f_{l}\right].
\label{eq:Euler equation main}
\end{eqnarray}
\end{widetext}

Here, $s$ is the entropy density, $w\equiv\mu n+Ts$ is the enthalpy
density and $\mathbf{J}^{n}$ is the particle current. The RHS of the equation yields momentum relaxation due to
the disorder collision integral. To linear order in $\mathbf{u}$,
the RHS equals $-\mathbf{u}/\bar{\tau}_{\parallel}^{\textrm{el}}$,
where $1/\bar{\tau}_{\parallel}^{\textrm{el}}$ describes the transport
elastic scattering rate off the impurities. 
The calculation of the elastic transport time  $\bar{\tau}_{\parallel}^{\textrm{el}}$
in terms of the microscopic parameters of our model is outlined in Appendix \ref{sec:Disorder-transport-scattering}. 
Note that in Eq. (\ref{eq:Euler equation main}) the electrochemical
field $\mathbf{E}-\mathbf{\nabla}\mu/e$ couples to
the charge density $n$ while the temperature gradient couples to
the entropy density $s$. At the charge neutrality point, the charge
density is zero while the entropy density is finite, making the zero-mode
couple effectively to the temperature gradient, but not to the electric
field. We note that in this work we do not consider viscosity, which would add a term proportional to $\nabla^2_{r} \mathbf{u}$ to Eq. (\ref{eq:Euler equation main}) and turn it to the Navier-Stokes equation. Viscosity establishes a Poiseuille flow profile near the boundary of the sample, up to a distance in the scale of the Gurzhi length $l_{G}\equiv\sqrt{\eta_0 \bar{\tau}_{\parallel}^{\textrm{el}} v_F^2 / w}$, with $\eta_0$ being the viscosity of the electron fluid \cite{Lucas2018}. Our approximation is thus valid when the sample width is much larger than the Gurzhi length.

\subsection{Thermoelectric linear response coefficients}

Next, we will derive the linear thermoelectric conductivities. The
total electric and energy currents are given by

\begin{equation}
\mathbf{J}^{c}=\sum_{l}\intop\left(d\mathbf{k}\right)\lambda_{l\mathbf{k}}^{c}f_{l\mathbf{k}}\mathbf{v}_{l\mathbf{k}}+\mathbf{J}_{\textrm{anomalous}}^{c},\label{eq:total current general}
\end{equation}

where we denote $c=e,Q$ for electric and thermal charges, with $\lambda_{l\mathbf{k}}^{e}=e,\lambda_{l\mathbf{k}}^{Q}=\epsilon_{l\mathbf{k}}-\mu$,
and $\mathbf{v}_{l\mathbf{k}}\equiv\partial\epsilon_{l\mathbf{k}}/\partial\mathbf{k}$
is the regular part of the velocity operator. While the first term
in Eq.  (\ref{eq:total current general}) will have contributions
only from the Fermi surface, the anomalous part includes Fermi sea
contributions and corresponds to the Berry curvature and magnetization
currents \cite{Cooper1997, Qin2011}, leading to transverse transport. As we will
argue, in the limit of linear dispersion and no tilt, transverse transport will
come solely from the anomalous term. We will first discuss the regular
part, and derive the longitudinal response coefficients.

We separate the regular part of the currents into ideal and
dissipative parts, writing

\begin{equation}
\sum_{l}\intop\left(d\mathbf{k}\right)\lambda_{l\mathbf{k}}^{c}f_{l\mathbf{k}}\mathbf{v}_{l\mathbf{k}} =\mathbf{J}_{\textrm{ideal }}^{c}+\mathbf{J}_{\textrm{diss.}}^{c}.
\end{equation}
These two parts correspond to the contributions from the local equilibrium
and the dissipative parts of the distribution function:

\begin{align}
\mathbf{J}_{\textrm{ideal }}^{c} & \equiv\sum_{l}\intop\left(d\mathbf{k}\right)\lambda_{l\mathbf{k}}^{c}n_{F}(x_{l\mathbf{k}})\mathbf{v}_{l\mathbf{k}},\\
\mathbf{J}_{\textrm{diss.}}^{c} & \equiv\sum_{l}\intop\left(d\mathbf{k}\right)\lambda_{l\mathbf{k}}^{c}\delta f_{l\mathbf{k}}\mathbf{v}_{l\mathbf{k}}.
\end{align}

The ideal part of the currents comes directly from the boost velocity.
For the electric current one finds

\begin{equation}
\mathbf{J}_{\textrm{ideal}}^{e}=en\mathbf{u},\label{eq:ideal e current}
\end{equation}

and for the heat current

\begin{equation}
\mathbf{J}_{\textrm{ideal}}^{Q}=Ts\mathbf{u}.\label{eq:ideal q current}
\end{equation}

The ideal currents describe the uniform motion of the electron fluid,
and they are limited only by the momentum relaxation mechanism, which
in our case is set by the disorder. By solving the Euler equation
[Eq. (\ref{eq:Euler equation main}), leading to the last equation
in Appendix \ref{sec:Euler equation derivation}] and using Eqs.
(\ref{eq:ideal e current}), (\ref{eq:ideal q current}) we find the
ideal parts of the currents in the presence of an electric field and
a temperature gradient.

The dissipative parts of the currents come from the part of the distribution
function that is not in local equilibrium. We calculate these by projecting
the electron-electron collision integral on the subspace orthogonal
to the zero-modes and approximating it to have one electron-electron
scattering time scale $\tau^{\textrm{e-e}}$ (see Appendix \ref{sec:Calculation-of-dissipative}
for details of the calculation). Importantly, the energy current $\mathbf{J}^{E}=\mathbf{J}^{Q}+\frac{\mu}{e}\mathbf{J}^{e}$
has no dissipative part. This is due to the linear dispersion relation,
which implies that the energy current is proportional to the momentum
density and thus is a conserved quantity of the electron-electron
collision integral [conversely, in the case of parabolic dispersion,
$\epsilon_{k}=k^{2}/(2m)$, the particle current is conserved and
has no dissipative part].

The linear thermoelectric response coefficients are defined by \cite{ashcroft1976solid}

\begin{align}
J_{\alpha}^{\textrm{e}} & =L_{\alpha\beta}^{\textrm{11}}\left(E_{\beta}-\frac{\nabla_{\beta}\mu}{e}\right)+L_{\alpha\beta}^{\textrm{12}}\left(-\nabla_{\beta}T\right),\label{eq:general Ls eq 1}\\
J_{\alpha}^{\textrm{Q}} & =L_{\alpha\beta}^{\textrm{21}}\left(E_{\beta}-\frac{\nabla_{\beta}\mu}{e}\right)+L_{\alpha\beta}^{\textrm{22}}\left(-\nabla_{\beta}T\right).\label{eq:general Ls eq 2}
\end{align}
It is clear that $L_{\alpha\beta}^{11} \equiv \sigma_{\alpha\beta}$ is the electric conductivity. Combining the ideal part of electric current [Eq. (\ref{eq:ideal e current})] with the dissipative part [Eqs. (\ref{dissipative_current1}), (\ref{dissipative_current2}) in the Appendix], we find 
the longitudinal electric conductivity to be given by
\begin{align}
L_{xx}^{11} & =\sigma_{xx}=\sigma_{xx}^{\textrm{ideal}}+\sigma_{xx}^{\textrm{diss.}},\\
\sigma_{xx}^{\textrm{\textrm{ideal}}} & =e^{2}v_{F}^{2}\frac{n^{2}}{w}\bar{\tau}_{\textrm{\ensuremath{\parallel}}}^{\textrm{el}},\\
\sigma_{xx}^{\textrm{\textrm{diss.}}} & =e^{2}v_{F}^{2}\left(\frac{1}{3}\frac{\partial n}{\partial\mu}-\frac{n^{2}}{w}\right)\tau^{\textrm{e-e}},
\end{align}
where we denoted the contributions from the ideal and dissipative parts of the current by $\sigma_{xx}^{\textrm{ideal}}$, $\sigma_{xx}^{\textrm{diss.}}$, respectively.
Since the condition for the hydrodynamic regime is $\tau^{\textrm{e-e}}\ll\bar{\tau}_{\textrm{\ensuremath{\parallel}}}^{\textrm{el}}$,
the ideal part of the conductivity is much larger than the dissipative
part, except when the chemical potential is in the very near vicinity
of the charge neutrality point, $\left|\mu/T\right|\ll\sqrt{\tau^{\textrm{e-e}}/\bar{\tau}_{\textrm{\ensuremath{\parallel}}}^{\textrm{el}}}$.
In the limit $\left|\mu/T\right| \gg 1$, the longitudinal conductivity
recovers the non-interacting value, which for the model of short-ranged
impurities is given by \cite{Sarma2015} \footnote{The right most expression in Eq. (\ref{eq:sigma_xx_infty}) has a factor two (the number of Weyl nodes) compared to the single node result, due to the assumption
of no internode scattering by the disorder. Therefore, the inverse of $\bar{\tau}_{\textrm{\ensuremath{\parallel}}}^{\textrm{el}}$ scales as the density of states
of a single node, while $\partial n/\partial\mu$ in the expression for $\sigma_{xx}$ gives the total density of states.}
\begin{equation}
\sigma_{xx}^{\infty}\equiv\sigma_{xx}(\left|\frac{\mu}{T}\right| \gg 1)=\left[e^{2}\frac{\partial n}{\partial\mu}v_{F}^{2}\bar{\tau}_{\textrm{\ensuremath{\parallel}}}^{\textrm{el}}/3\right]=\frac{e^{2}v_{F}^{2}}{\pi\gamma}.\label{eq:sigma_xx_infty}
\end{equation}

Similarly, we find for the rest of the longitudinal thermoelectric
response coefficients

\begin{align}
L_{xx}^{\textrm{21}} & =TL_{xx}^{\textrm{12}}=\frac{T}{e}\left[\frac{s}{n}\sigma_{xx}^{\textrm{ideal}}-\frac{\mu}{T}\sigma_{xx}^{\textrm{diss.}}\right],\label{eq:L12 xx}\\
L_{xx}^{\textrm{22}} & =\frac{T}{e^{2}}\left[\left(\frac{s}{n}\right)^{2}\sigma_{xx}^{\textrm{ideal}}+\left(\frac{\mu}{T}\right)^{2}\sigma_{xx}^{\textrm{diss.}}\right].
\end{align}

In a WSM, the Fermi surface parts of the currents can give additional contributions
to the transverse response coefficients, via processes such as skew scattering
and side jumps \cite{Sinitsyn2007, Ado2015}. In the Boltzmann equation formalism,
the velocity operator and collision integral acquire corrections, yielding
contributions to the anomalous Hall conductivities from the first
term in Eq. (\ref{eq:total current general}). However, since we are
considering the vicinity of the Weyl nodes and in the absence of a tilt, 
the low-energy Hamiltonian for each node [Eq. (\ref{eq:Hamiltonian})] is time-reversal symmetric and
the Fermi surface contributions to the anomalous Hall conductivities vanish. Therefore
the problem is simplified, and the transverse response will be solely
from Fermi sea terms, making up the terms $\mathbf{J}_{\textrm{anomalous}}^{c}$
which we will describe now. Alternatively, the anomalous currents can be viewed
as contributions from the Fermi arc surface states in a finite sample.
Due to their Fermi sea nature, the transverse conductivities will not
be affected by the presence of electron-electron
collisions, and will be equivalent to the non-interacting intrinsic
anomalous Hall conductivities \cite{Xiao2010, Qin2011}.

When electrons with Berry curvature are accelerated, the velocity
operator acquires an anomalous contribution,

\begin{equation}
\dot{\mathbf{r}}=\frac{\partial\epsilon_{l\mathbf{k}}}{\partial\mathbf{k}}+\dot{\mathbf{k}}\times\mathbf{\Omega}_{l\mathbf{k}}.
\end{equation}

Here, $\mathbf{\Omega}_{l\mathbf{k}}$ denotes the
Berry curvature. This Berry curvature contribution gives rise to the
anomalous Hall conductivity,

\begin{equation}
L_{xy}^{\textrm{11}}=\sigma_{xy}=e^{2}\sum_{l}\intop\left(d^{3}k\right)f_{l\mathbf{k}}^{0}\left(\Omega_{l\mathbf{k}}\right)_{z},\label{eq:sigma_xy general}
\end{equation}
where $f_{l\mathbf{k}}^{0}=n_{F}\left[(\epsilon_{l\mathbf{k}}-\mu)/T\right]$
is the unperturbed (global) equilibrium distribution function. Additionally,
due to magnetization currents, the presence of the electrochemical
potential and temperature gradients gives rise to contributions to
the transverse electric and thermal currents \cite{Cooper1997}. 
With these contributions, the transverse electro-thermal and thermal-thermal responses can be written in the following form \cite{Smrcka1977},
which ensures the validity of the Mott relations and the Wiedemann-Franz law in the low temperature-limit:

\begin{align}
L_{xy}^{\textrm{21}} & =TL_{xy}^{\textrm{12}}=\frac{T}{e}\intop d\epsilon\frac{\epsilon-\mu}{T}\sigma_{xy}^{\textrm{}}(\epsilon)\left(-\frac{\partial f^{0}}{\partial\epsilon}\right),\label{eq:L21_xy_general}\\
L_{xy}^{\textrm{22}} & =\frac{T}{e^{2}}\intop d\epsilon\frac{\left(\epsilon-\mu\right)^{2}}{T^{2}}\sigma_{xy}^{\textrm{}}(\epsilon)\left(-\frac{\partial f^{0}}{\partial\epsilon}\right)\label{eq:L22_xy general}.
\end{align}

Here, $\sigma_{xy}(\epsilon)$ is the anomalous Hall conductivity
at zero temperature and chemical potential $\mu=\epsilon$.
At the charge neutrality point, the anomalous Hall conductivity {[Eq. (\ref{eq:sigma_xy general})]}
is proportional to the distance between the Weyl nodes \cite{Burkov2011b},

\begin{equation}
\sigma_{xy}(\text{\ensuremath{\epsilon=0)}}=\frac{e^{2}}{4\pi^{2}}\Delta_k.\label{eq:sigma_xy burkov result}
\end{equation}

In a Weyl semimetal, $\sigma_{xy}(\epsilon)$ varies over an energy
scale of order $\epsilon \sim \Delta_k v_{F}$, the energy in which the two separate
Fermi surfaces surrounding each node merge through Lifshitz transition
\cite{Burkov2014}. Consistently with the assumption of being near
the Weyl nodes, we take the AHE conductivity to be constant, $\sigma_{xy}(\epsilon)=\sigma_{xy}(\epsilon=0$),
neglecting subleading terms in $\mu/\left(\Delta_k v_{F}\right)$ and $T/\left(\Delta_k v_{F}\right)$.
Then, the rest of the transverse response coefficients are
implied by Eqs. (\ref{eq:L21_xy_general}), (\ref{eq:L22_xy general})
to be given by

\begin{align}
L_{xy}^{\textrm{21}} & =TL_{xy}^{\textrm{12}}=0,\label{eq:L_12_xy}\\
L_{xy}^{\textrm{22}} & =\frac{\pi^{2}T}{3e^{2}}\sigma_{xy}^{\textrm{}}. \label{eq:L_xy_22 intrinsic}
\end{align}

We emphasize that the above formulas hold in the vicinity of Weyl
nodes ($\left| \mu \right|,T\ll \Delta_k v_{F}$) where the dispersion is linear and in the absence of a tilt. Relaxing either of
these assumptions breaks the TRS of the single-node Hamiltonian. Breaking the TRS of a single node introduces Fermi surface contributions to the Hall conductivities.
In the vicinity of the Weyl nodes, these Fermi surface contributions to the electric and thermal Hall conductivities are expected to be subleading in
$\max[\left| \mu \right|,T ] / \left(\Delta_k v_{F} \right)$ compared to the intrinsic contributions given in Eqs.
(\ref{eq:sigma_xy burkov result}), (\ref{eq:L_xy_22 intrinsic}) \cite{Burkov2014}. The Fermi surface contributions
should include contributions due to electron-electron scattering
\cite{Pesin2018, Glazov2022}, as well as the electron-impurity scattering contributions known for
non-interacting systems \cite{Sinitsyn2007, Ado2015, Steiner2017, zhang2023disorder}.
The presence of a tilt would give finite contributions to the thermoelectric coefficients $L_{xy}^{\textrm{12}}, L_{xy}^{\textrm{21}}$
due to the breaking of particle-hole symmetry \cite{Ferreiros2017, Saha2018}, even close to the Weyl nodes.
The above discussion excludes the case of a strong tilt, $u_t/v_F>1$ (leading to a type-II WSM \cite{Armitage2018}),
in which the value of $\sigma_{xy}$ at the neutrality point is also modified \cite{Zyuzina2016}.
The quantitative analysis of these regimes is outside the scope of our work.

To summarize, the thermoelectric responses are composed of three different modes
of transport: momentum density zero-mode, relaxed by disorder;
dissipative mode, relaxed by electron-electron
interactions; and anomalous Hall part stemming from the topological
band structure. Next, we will explore the interplay of these three
mechanisms on the Seebeck response and heat conductivity.

\subsection{Seebeck tensor and Heat conductivity}

The current responses in an experimental setup depend on the boundary
conditions. The response coefficients $L_{\alpha\beta}^{ij}$ defined
in Eqs. (\ref{eq:general Ls eq 1}), (\ref{eq:general Ls eq 2}) correspond
to applying either an electric field or a temperature gradient while
keeping the other zero. For thermal conductivity, often the experimental
setup is an open circuit, where the electric field is not controlled
but rather the electric current is forced to be zero. In this setup
one measures the Seebeck response, the electrochemical potential response
to a temperature gradient, defined by

\begin{equation}
S_{\alpha\beta}\equiv\left[\frac{E_{\alpha}-\nabla_{\alpha}\frac{\mu}{e}}{\nabla_{\beta}T}\right]_{\mathbf{J}^{\textrm{e}}=0},
\end{equation}

and the open circuit heat conductivity,

\begin{equation}
\kappa_{\alpha\beta}\equiv\left[-\frac{J_{\alpha}^{\textrm{q}}}{\nabla_{\beta}T}\right]_{\mathbf{J}^{\textrm{e}}=0}.
\end{equation}

By using the responses in Eqs. (\ref{eq:general Ls eq 1}), (\ref{eq:general Ls eq 2})
and setting $\mathbf{J}^{\textrm{e}}=0$ one obtains

\begin{align}
S_{\alpha\beta} & =\rho_{\alpha\gamma}L_{\gamma\beta}^{\textrm{12}},\\
\kappa_{\alpha\beta} & =L_{\alpha\beta}^{\textrm{22}}-L_{\alpha\gamma}^{\textrm{21}}S_{\gamma\beta},
\end{align}

where $\rho_{\alpha\beta}\equiv\left(\sigma^{-1}\right)_{\alpha\beta}$
is the resistivity (note that this involves inverting a matrix in
x-y space). In the hydrodynamic regime, the ideal part of the
thermoelectric response [first term in Eq. (\ref{eq:L12 xx})] dominates by a factor
of $\tau^{\textrm{el}}/\tau^{\textrm{e-e}}$, and we get

\begin{align}
S_{xx} & =\frac{s}{en}\frac{\sigma_{xx}\sigma_{xx}^{\textrm{ideal}}}{\sigma_{xx}^{2}+\sigma_{xy}^{2}},\label{eq:Sxx}\\
S_{xy} & =-\frac{s}{en}\frac{\sigma_{xy}\sigma_{xx}^{\textrm{ideal}}}{\sigma_{xx}^{2}+\sigma_{xy}^{2}},\label{eq:Sxy}
\end{align}

for the Seebeck tensor, and the following expressions for the heat
conductivity:

\begin{widetext}
\begin{align}
\kappa_{xx} & =\frac{T}{e^{2}\left(\sigma_{xx}^{2}+\sigma_{xy}^{2}\right)}\left\{ \sigma_{xx}^{\textrm{diss.}}\left[\left(\frac{s}{n}+\frac{\mu}{T}\right)^{2}\sigma_{xx}\sigma_{xx}^{\textrm{ideal}}+\left(\frac{\mu}{T}\right)^{2}\left(\sigma_{xy}^{\textrm{}}\right)^{2}\right]+\left(\frac{s}{n}\right)^{2}\sigma_{xx}^{\textrm{ideal}}\left(\sigma_{xy}^{\textrm{}}\right)^{2}\right\} ,\label{eq:kappa_xx full}\\
\kappa_{xy} & =\frac{T}{e^{2}\left(\sigma_{xx}^{2}+\sigma_{xy}^{2}\right)}\sigma_{xy}^{\textrm{}}\left[\frac{\pi^{2}}{3}\left(\sigma_{xx}^{2}+\sigma_{xy}^{2}\right)+\left(\frac{s}{n}\sigma_{xx}^{\textrm{ideal}}-\frac{\mu}{T}\sigma_{xx}^{\textrm{diss.}}\right)^{2}\right].\label{eq:kappa_xy full}
\end{align}
\end{widetext}

\section{Discussion of the results}
Now we discuss the results Eqs. (\ref{eq:Sxx})-(\ref{eq:kappa_xy full}) and their asymptotic behavior.
Specifically, we will consider the behavior as a function of the anomalous
Hall angle (AHA) $\tan \Theta^\textrm{H} = \sigma_{xy}/\sigma_{xx}^{\infty}$.
Let us first elaborate on the AHA in our model and in possible physical systems.
In our model, $\sigma_{xy}$ and $\sigma_{xx}^{\infty}$ [Eqs. (\ref{eq:sigma_xy burkov result}), (\ref{eq:sigma_xx_infty})]
can be tuned independently, the first being controlled
by the distance between the Weyl nodes $\Delta_k$ and the latter
by the disorder transport time $\bar{\tau}_{\parallel}^{\textrm{el}}$.
Experimentally, while typical values of the AHA are of a few percent, recent advances
have allowed to achieve AHA values of 0.21 at room temperature and 0.33
at $T=2\textrm{K}$ in WSM candidates \cite{Li2020, Chen2021}.
Foreseeing further development, we include in our analysis
also the possibility of AHA values greater than one.

\subsection{Seebeck coefficients}

The Seebeck coefficients [Eqs. (\ref{eq:Sxx}), (\ref{eq:Sxy})] quantify
the ratio between the thermoelectric response coefficients $L_{\alpha\beta}^{\textrm{12}}$
and the electric conductivities. When the electric conductivity is
also dominated by the local equilibrium part, the longitudinal Seebeck coefficient
$S_{xx}$ reaches the hydrodynamic limit of entropy per electric charge,
$s/\left(ne\right)$ \cite{Lucas2018}. Let us note that the limit
of entropy per electric charge is universal for various types of metals
in the limit $\left|\mu/T\right|\gg1$ \cite{Behnia2004}, giving a small Seebeck
response for normal metals. In the hydrodynamic case, this limit persists
in the entropy-dominated Dirac-fluid regime ($\left|\mu/T\right|\ll1$) in which
$s/\left(ne\right)\gg1$, allowing a large Seebeck response (see Appendix \ref{sec:Seebeck-coefficient-for}
for comparison with the non-interacting case).

\begin{figure}[t]
\begin{centering}
\includegraphics[scale=0.445]{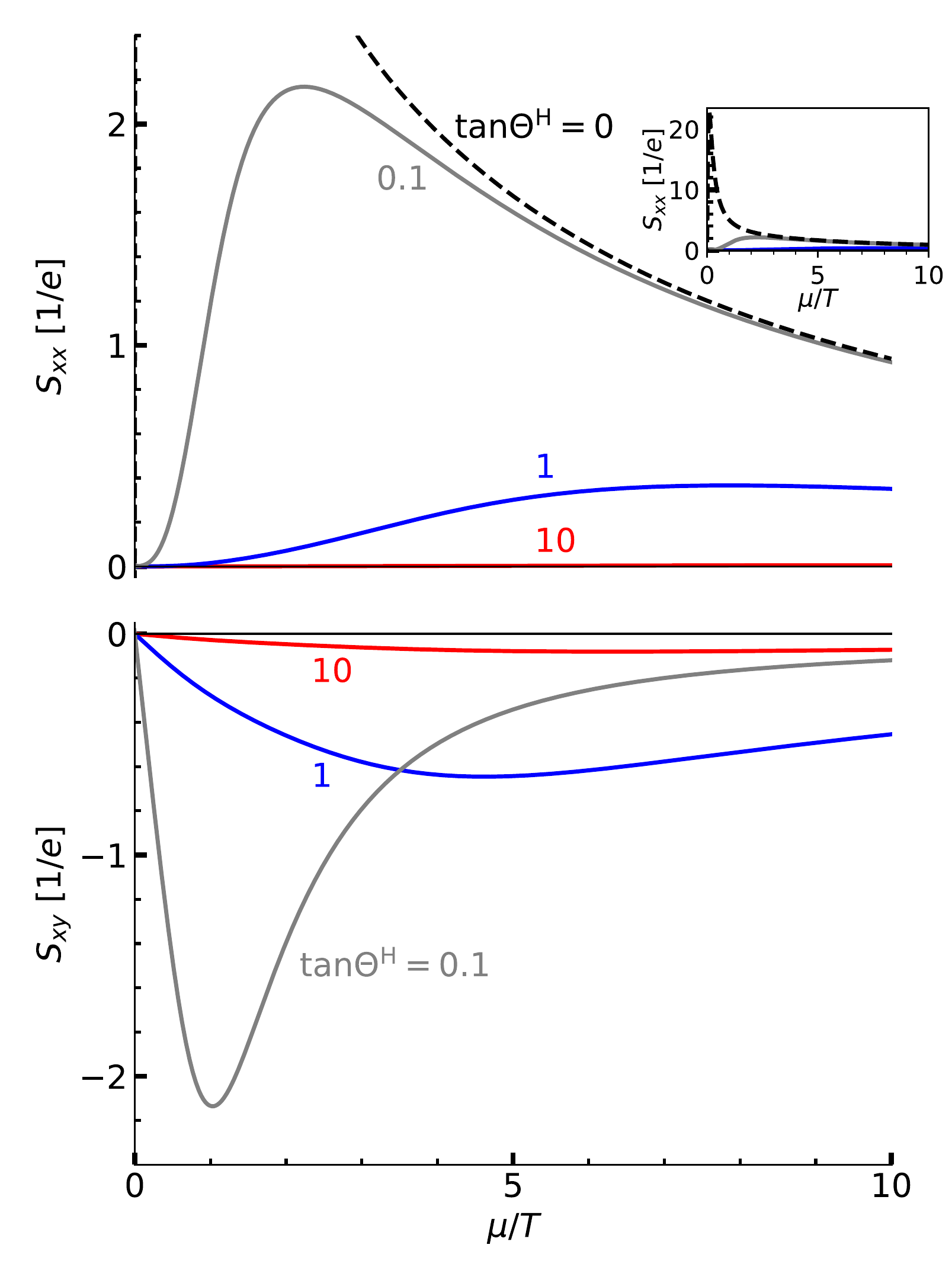}
\par\end{centering}
\caption{Longitudinal (top) and transverse (bottom) Seebeck coefficients for varying chemical potential and $\tan \Theta^\textrm{H}=\sigma_{xy}/\sigma_{xx}^{\infty}$ values. Inset: zoomed out view of the longitudinal Seebeck coefficient. For $\tan \Theta^{\textrm{H}}=0$, $S_{xx}\simeq\sqrt{\sigma_{xx}^{\infty}/\sigma_{xx}^{\textrm{diss.}}}/e\sim1/\sqrt{T}$ near the neutrality point.
	\label{fig:Seebeck}}
\end{figure}

In the absence of the AHE, $S_{xx}$ goes below the hydrodynamic limit
only for very small carrier density, where the dissipative part of
the electric conductivity dominates. With a finite anomalous Hall
conductivity, $\rho_{xx}$ is increased and $S_{xx}$ is suppressed
at larger range of carrier densities (Fig. \ref{fig:Seebeck}, top). On the
other hand, the AHE enables a transverse Seebeck effect (the anomalous
Nernst effect) via the combination of transverse resistivity and the
longitudinal thermoelectric response (Fig. \ref{fig:Seebeck}, bottom). Interestingly,
for small values of $\sigma_{xy}$, the longitudinal and transverse
Seebeck coefficients reach the same maximum value, which we find to
be

\begin{eqnarray}&&
\left(S_{xx}\right)_{\textrm{max}}=\left(\left|S_{xy}\right|\right)_{\textrm{max}} \nonumber \\&&
\underset{\left[\sigma_{xy}^{\textrm{}}/\sigma_{xx}^{\infty}\ll1\right]}{=}\frac{7\pi}{20}\sqrt{\frac{21}{31\cdot3^{1/2}}}\sqrt{\frac{\sigma_{xx}^{\infty}}{\sigma_{xy}^{\textrm{}}}}\frac{1}{e}.\label{eq:Sxx Sxy max}
\end{eqnarray}
These maxima are reached at chemical potential values where $\sigma_{xx}\simeq\sigma_{xy}$.
This resembles the maximum power transfer theorem in electric circuits,
stating that maximum power is delivered to a load when the resistances
of the load and the power source are equal \cite{Floyd1997}. The
increase in the maximum of $S_{xy}$ as the AHE is becoming weaker
(i.e. as we decrease the distance between the Weyl nodes) is limited
to the point where the anomalous Hall conductivity becomes comparable
with the dissipative part of the conductivity, $\sigma_{xy}\simeq\sigma_{xx}^{\textrm{diss.}}$.
Lowering $\sigma_{xy}$ further eliminates the transverse Seebeck
response, as expected when time-reversal symmetry is restored, and
brings the maximal value for $S_{xx}$ to be given by Eq. (\ref{eq:Sxx Sxy max})
with the replacement $\sigma_{xy}\rightarrow\sigma_{xx}^{\textrm{diss.}}/3$.

\subsection{Longitudinal heat conductivity}

Next, we analyze the result for the longitudinal heat conductivity,
Eq. (\ref{eq:kappa_xx full}). In the Dirac fluid regime, the entropy
terms dominate the heat conductivity and we find

\begin{equation}
\kappa_{xx}(\left|\mu/T\right|\ll1)=\frac{T}{e^{2}}\left(\frac{s}{n}\right)^{2}\sigma_{xx}^{\textrm{ideal}}\left[1-\rho_{xx}\sigma_{xx}^{\textrm{ideal}}\right].\label{eq:kappa_xx eq}
\end{equation}

The most efficient energy transport is via the momentum density zero-mode. At the charge
neutrality point, this mode transports energy but no electric
current. The electron-electron collisions are unable to relax the
energy current, and the heat conductivity behaves similarly as in a non-interacting
system with the same disorder. In the limit $\mu/T\rightarrow0$ we
find the value $\kappa_{xx}\rightarrow\frac{343\pi^{2}}{2325}\frac{T}{e^{2}}\sigma_{xx}^{\infty}$,
similar to the result in 2D graphene \cite{Lucas2018}. This can be
written by a Drude-like formula,

\begin{equation}
\kappa_{xx}(\mu/T=0)=c_{v}\frac{v_{F}^{2}\bar{\tau}_{\parallel}^{\textrm{el}}}{3},
\end{equation}
with $c_{v}=T\partial s / \partial T$ being the heat capacity
per volume. We note that in the model of short-ranged disorder, the
electron-electron collisions do affect $\kappa_{xx}(\mu/T=0)$ indirectly
by modifying the effective elastic scattering time. Due to the e-e
collisions, the elastic scattering rate $1/\tau_{\parallel}^{\textrm{el}}(\epsilon)$
is averaged over an energy window with the width of the temperature,
causing $\kappa_{xx}(\mu/T=0)$ to be larger by a numerical factor
($\approx30$) compared to the non-interacting result (see Appendix
\ref{sec:Disorder-transport-scattering}). More dramatically, as
the electron-electron collisions cause $\sigma_{xx}$ to be governed
by the smaller time scale $\tau^{\textrm{e-e}}$, they greatly enhance
the Lorenz ratio $\mathcal{L}_{xx}\equiv\kappa_{xx}/\left(T\sigma_{xx}\right)$
[which for non-interacting electrons is given by $\mathcal{L}_{0}=\pi^{2}/\left(3e^{2}\right)$]
near the charge neutrality point. At the charge neutrality point,
the Lorenz ratio is given by

\begin{equation}
\mathcal{L}_{xx}(\mu/T=0)=\frac{7\pi^{2}}{5e^{2}}\frac{\bar{\tau}_{\parallel}^{\textrm{el}}}{\tau^{\textrm{e-e}}}.
\end{equation}

\begin{figure}[t]
\begin{centering}
\includegraphics[scale=0.45]{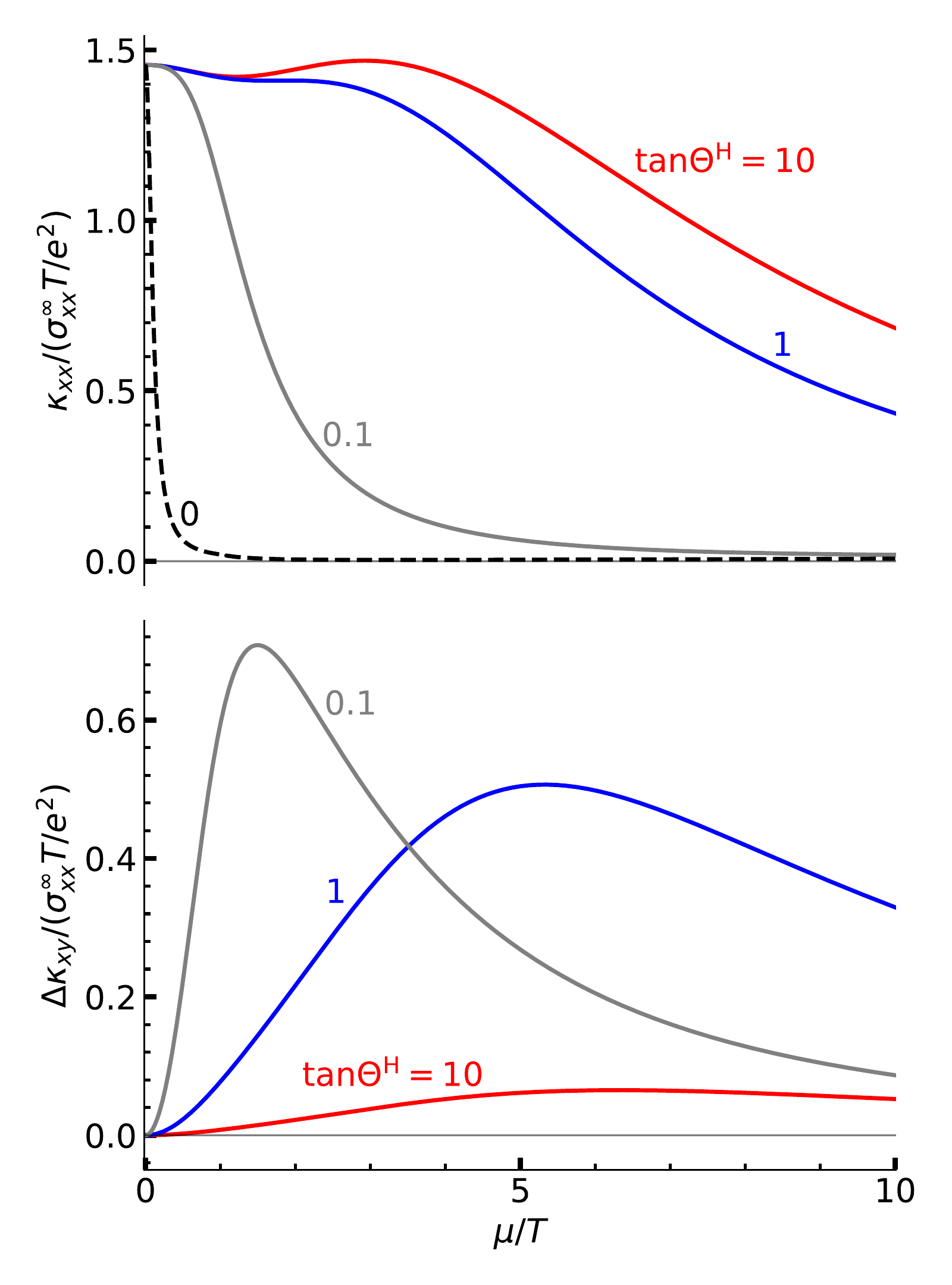}
\par\end{centering}
\caption{Longitudinal heat conductivity $\kappa_{xx}$ (top) and deviation of the transverse heat conductivity from the WF value $\Delta\kappa_{xy}\equiv\kappa_{xy}-\mathcal{L}_{0}T\sigma_{xy}$ (bottom), for varying chemical potential and $\tan \Theta^\textrm{H}=\sigma_{xy}/\sigma_{xx}^{\infty}$
 values. \label{fig:kappa}}
\end{figure}

At charge neutrality, the longitudinal heat conductivity is unaffected
by the AHE. This differs when the chemical potential is increased.
Then, the momentum density zero-mode begins to carry also particle current as well
as energy current, ultimately rendering it unable to transfer heat
without convection. This corresponds to a cancellation by the longitudinal
Seebeck response $S_{xx}$, which is the second term in Eq. (\ref{eq:kappa_xx eq}).
However, a finite Hall conductivity decreases $S_{xx}$ and thus broadens
the range in which $\kappa_{xx}$ is not small, as can be seen in
Fig. \ref{fig:kappa}. In the limit of $\sigma_{xy}\gg\sigma_{xx}^{\infty}$,
this leads to a second peak in $\kappa_{xx}$ at the region between
the Dirac fluid and Fermi liquid regimes, which is slightly higher
than the value at the charge neutrality point, $\kappa_{xx}(\left|\mu/T\right|\approx2.93)\approx1.467\frac{T}{e^{2}}\sigma_{xx}^{\infty}$.

Eventually, for $\left|\mu/T\right|\gg1$  (we remind that throughout this work, we limit ourselves to being far from the Lifshitz transition, $\left| \mu \right| \ll v_F \Delta_k$)
the thermal conductivity becomes dominated
by the dissipative part of the current, decaying to $\kappa_{xx}(\left|\mu/T\right|\gg1)=\frac{\mu}{e^{2}}\frac{\mu}{T}\sigma_{xx}^{\textrm{diss.}}$.
This can be written as

\begin{equation}
\kappa_{xx}(\left|\mu/T\right|\gg1)=c_{v}\frac{v_{F}^{2}\tau^{\textrm{e-e}}}{3}.
\end{equation}

This gives a parameterically small Lorenz ratio,

\begin{equation}
\mathcal{L}_{xx}=\mathcal{L}_{0}\frac
{\tau^{\textrm{e-e}}}
{\bar{\tau}_{\parallel}^{\textrm{el}}}
,\label{eq:Lorenz_xx_degenerate}
\end{equation}

reflecting that charge and thermal transport are limited by two different
scattering rates \cite{Principi2015}.  Suppression of the Lorenz ratio has
been observed in ultra-clean samples of WP$_2$ and Sb \cite{Gooth2018, Jaoui2018, Jaoui2021},  
both materials being in the degenerate regime. Nonetheless, this suppression does not necessarily
arise from the materials being in the hydrodynamic regime; see Appendix \ref{sec:Experimental} for further discussion.

\subsection{Transverse heat conductivity}

We now turn to analyze the transverse heat conductivity, Eq. (\ref{eq:kappa_xy full}).
Keeping the leading terms we find

\begin{equation}
\kappa_{xy}=\frac{\pi^{2}T}{3e^{2}}\sigma_{xy}-\frac{Ts}{en}\sigma_{xx}^{\textrm{ideal}}S_{xy}.
\end{equation}

The first term corresponds to the Wiedemann-Franz result in the non-interacting
regime, for which the Lorenz ratio is $\mathcal{L}_{xy}=\mathcal{L}_{0}$.
The second term accounts for heat transfer by the momentum density zero-mode, which acquires a transverse component due to the transverse Seebeck response. Following
the behavior of the transverse Seebeck coefficient $S_{xy}$, this
additional term $\Delta\kappa_{xy}\equiv\kappa_{xy}-\mathcal{L}_{0}T\sigma_{xy}$
is appreciable for small Hall-to-longitudinal conductivity ratios
$\sigma_{xy}^{\textrm{}}/\sigma_{xx}^{\infty}$ (Fig. \ref{fig:kappa}, bottom).
In the limit $\sigma_{xy}^{\textrm{}}/\sigma_{xx}^{\infty}\ll1$ (but
$\sigma_{xy}^{\textrm{}}\gg\sigma_{xx}^{\textrm{diss.}}$), the deviation
from Wiedemann-Franz law $\Delta\kappa_{xy}$ reaches a maximum

\begin{equation}
\left(\Delta\kappa_{xy}\right)_{\textrm{max}}\underset{\left[\sigma_{xy}^{\textrm{}}/\sigma_{xx}^{\infty}\ll1\right]}{=}\frac{343\pi^{2}}{4650}\frac{T}{e^{2}}\sigma_{xx}^{\infty},
\end{equation}

which occurs at the carrier density for which $\sigma_{xx}^{\textrm{ideal}}=\sigma_{xy}^{\textrm{}}$.
Note that this peak is much higher than $\kappa_{xy}(\mu/T=0)$ in
this limit, leading to a large Lorenz ratio $\mathcal{L}_{xy}=\kappa_{xy}/\left(T\sigma_{xy}\right)\simeq \mathcal{L}_0 \sigma_{xx}^{\infty}/\sigma_{xy}$.
For large $\sigma_{xy}/\sigma_{xx}^{\infty}$ ratios, the transverse
Seebeck response is suppressed, and deviation from the Wiedemann-Franz
law reaches a maximal value of

\begin{equation}
\Delta\kappa_{xy}(\left|\mu/T\right|\approx6.34)\underset{\left[\sigma_{xy}^{\textrm{}}/\sigma_{xx}^{\infty}\gg1\right]}{=}0.654\frac{T}{e^{2}}\frac{\left(\sigma_{xx}^{\infty}\right)^{2}}{\sigma_{xy}^{\textrm{}}}.
\end{equation}

Going further into the Fermi liquid regime $\left|\mu/T\right|\gg1$, the transverse
Seebeck response diminishes and the transverse thermal conductivity
decays back to the non-interacting Wiedemann-Franz result.
%

\section{Conclusions and outlook}

We considered a pair of TRS-breaking Weyl nodes with strong electron-electron
scattering. We derived the hydrodynamic equations and computed the
thermoelectric response coefficients. The electric and thermal currents
are carried via three mechanisms: long-lived hydrodynamic zero-mode
limited by disorder, dissipative modes relaxed by electron-electron
collisions, and transverse transport due to the Berry curvature and
magnetization currents.

We computed the effect of time-reversal symmetry breaking on the Seebeck
response and heat conductivities. The anomalous Hall conductivity
suppresses the longitudinal Seebeck response, which otherwise reaches
the hydrodynamic limit of entropy per electric charge, but enables
a transverse Seebeck response. The transverse Seebeck response is
maximal when the anomalous Hall conductivity is of intermediate strength,
smaller than the non-interacting limit of the longitudinal conductivity
$\sigma_{xx}^{\infty}$, but larger than the electron-electron collisions
limited conductivity $\sigma_{xx}^{\textrm{diss.}}$.

Due to its effect on the Seebeck response, the anomalous Hall conductivity
enhances the longitudinal thermal conductivity $\kappa_{xx}$. In
the TRS case, the longitudinal heat conductivity in the limiting cases
$\left|\mu/T\right|\ll1$ and $\left|\mu/T\right|\gg1$ can be written with a Drude-like formula,
$\kappa_{xx}=c_{v} v_{F}^{2}\tau/3$, with $\tau=\bar{\tau}_{\parallel}^{\textrm{el}}$
for $\left|\mu/T\right|\ll1$ and $\tau=\tau^{\textrm{e-e}}$ for $\left|\mu/T\right|\gg1$,
describing heat transport via the momentum density zero-mode and dissipative modes,
respectively. Breaking TRS, the anomalous Hall conductivity enables
the longer lived zero-mode to conduct heat up until the range of $\left|\mu/T\right|\sim1$,
enhancing $\kappa_{xx}$.

The transverse heat conductivity $\kappa_{xy}$ behaves according to
the Wiedemann-Franz law at the charge neutrality point. 
Away from the neutrality point the Wiedemann-Franz law is violated due to the large Seebeck effect.
This violation is maximal for intermediate TRS-breaking, where the anomalous Hall
conductivity $\sigma_{xy}$ is between the two limits of the longitudinal
conductivity, $\sigma_{xx}^{\textrm{diss.}}$ and $\sigma_{xx}^{\infty}$.

In our work we assumed $\tau^\textrm{e-e}$ to be the shortest scattering rate in the system.
This imposes a specific temperature window, where electron-phonon scattering can be neglected,
as well as requiring sufficient purity of the sample. Although this parameter range
is plausible, in current experiments a clear separation of scales is not yet fully achieved.
In Appendix \ref{sec:Experimental}  we briefly review the typical  scales for scattering mechanisms  in 
materials that are in (or close to) the hydrodynamic regime.
In this work we focused on TRS-breaking WSMs. These materials are typically more disordered,
and experimental realizations do not yet reach the required purity for hydrodynamics.
A viable path towards the hydrodynamic regime in TRS-breaking WSMs  is likely 
to be via thin films. Such films have been produced recently and have less disorder
and consequently smaller residual resistivity compared to bulk samples \cite{Tanaka2020}.

Finally, we compare the heat transport in a TRS-breaking WSM with that of a relativistic fluid in an external magnetic field \cite{Hartnoll2007, Muller2008, Miiller2009} and stress the main differences. 
The anomalous Hall currents which flow in a WSM are additive to the hydrodynamic flow, and do not affect the longitudinal electric and thermoelectric responses. This is in contrast to the case of a magnetic field, which rotates the boost velocity, lowering the longitudinal thermoelectric conductivities $L_{xx}^{ij}$ while giving rise to  ideal contributions to the transverse conductivities $L_{xy}^{ij}$.
This difference has several implications.

The longitudinal thermal conductivity  $\kappa_{xx}$ in a WSM is enhanced with the increase of the anomalous Hall conductivity  $\sigma_{xy}$. Conversely, an external magnetic field can only decrease the longitudinal thermal conductivity.  The transverse Lorenz ratio in WSM in the regime of intermediate AHE strength, $\sigma_{xx}^{\textrm{diss.}} \ll \sigma_{xy} \ll \sigma_{xx}^{\infty}$, reaches a maximum of $\mathcal{L}_{xy} \sim \sigma_{xx}^{\infty}/\sigma_{xy}$. For graphene in a weak magnetic field, the transverse Lorenz ratio can reach a maximum of $\mathcal{L}_{xy} \sim \bar{\tau}_{\parallel}^\textrm{el}/\tau^\textrm{e-e}$. As we see, the transverse Lorenz ratios at the maxima in these two systems are parametrically different.

We end this section with a brief discussion of future directions.
In this work, we have limited our study to a WSM in the vicinity of the neutrality point.
As one deviates from the neutrality point, the Dirac cones acquire curvature and the roles played 
by the various zero-modes change.
In addition, a finite Fermi surface contribution to the transverse thermoelectric response
coefficients $L_{xy}^{ij}$ appears.
In particular, electron-electron scattering induces side jumps and skew scattering \cite{Pesin2018, Glazov2022},
similar to the disorder-induced processes which contribute to
the non-interacting anomalous Hall conductivity \cite{Sinitsyn2007, Ado2015}. It is interesting to see
how such processes affect thermoelectric transport in the hydrodynamic
regime.  We plan to  address these questions in future work.


\begin{acknowledgments}

The authors are grateful to  B. Yan, T. Holder, D. Kaplan,  K. Michaeli, A.D. Mirlin, M. M{\"u}ller and M. Sch{\"u}tt for useful discussions.
This research was supported by ISF-China 3119/19 and ISF 1355/20. Y. M. thanks the PhD scholarship of the Israeli Scholarship
Education Foundation (ISEF) for excellence in academic and social leadership.
\end{acknowledgments}

\appendix

\section{Derivation of the Euler equation\label{sec:Euler equation derivation}}

Here we present a derivation of the hydrodynamic equations for linearly
dispersing Weyl nodes, leading to the Euler equation. Starting from
the Boltzmann equation [Eq. (\ref{eq:Boltzmann equation})], one derives
conservation equations by multiplying by conserved quantities of the
electron-electron collision integral and integrating over all states.
Here, these conserved quantities are the number, momentum and energy
densities. In this three-mode treatment, we are summing over both Weyl nodes,
utilizing our assumption that the nodes are related by inversion symmetry.
Multiplying the Boltzmann equation by the corresponding
charges $\lambda_{l\mathbf{k}}^{c}\in\left\{ 1,\mathbf{k},\epsilon_{l\mathbf{k}}\right\} $
and integrating over all states, we find

\begin{widetext}
\begin{align}
\frac{\partial n}{\partial t}+\mathbf{\nabla}_{r}\cdot\mathbf{J}^{n} & =0,\\
\frac{\partial\pi^{i}}{\partial t}+\frac{\partial\Pi_{E}^{ij}}{\partial r^{j}}-eE^{i}n & =\sum_{l}\intop\left(d^{3}k\right)k^{i}I_{\mathrm{ imp}}\left[f_{l}(t,\mathbf{r},\mathbf{k})\right],\label{eq:momentum conservation}\\
\frac{\partial n_{E}}{\partial t}+\mathbf{\nabla}_{r}\cdot\mathbf{J}^{E}-e\mathbf{E}\cdot\mathbf{J}^{n} & =0,\label{eq:energy conservation}
\end{align}
\end{widetext}

where the number, momentum and energy densities $n^{c}\in\left\{ n,\mathbf{\pi},n_{E}\right\}$ are
given by

\begin{equation}
n^{c}(\mathbf{r},t)=\sum_{l}\intop\left(d^{3}k\right)\lambda_{l \mathbf{k}}^{c}f_{l \mathbf{k}}(\mathbf{r},t),
\end{equation}

$\Pi_{E}^{ij}=\sum_{l}\intop\left(d^{3}k\right)k^{i} v_{l \mathbf{k}}^{j}f_{l \mathbf{k}}$
is the momentum-flux tensor and $\mathbf{J}^{n},\mathbf{J}^{E}$
are the particle and energy currents. The densities $n,\mathbf{\pi},n_{\epsilon}$
correspond to their conjugate potentials (chemical potential, boost
velocity and temperature) which parametrize the zero-mode distribution
function. In the equation for momentum conservation, Eq. (\ref{eq:momentum conservation}), we assumed equal chemical potential for pairing Weyl nodes, due to the absence of a magnetic field. If we relax this assumption and introduce chiral charge, Eq. (\ref{eq:momentum conservation}) acquires a term due originating from the anomalous velocity \cite{Gorbar2018}.

Calculation of the momentum density, energy current density and momentum-flux
tensor yields
\begin{align}
\mathbf{\pi} & =\frac{w}{v_{F}^{2}}\mathbf{u},\\
\mathbf{J}^{E} & =w\mathbf{u},\\
\Pi_{E}^{ij} & =p\delta^{ij}+w\frac{u^{i}u^{j}}{v_{F}^{2}} + \delta \Pi_{E}^{ij}.
\end{align}

Here, $p$ is the pressure and $w=n_{\epsilon}+p$ is the enthalpy
density. $\delta \Pi_{E}^{ij}$ is a dissipative correction to the momentum-flux tensor which corresponds to viscosity, and is proportional to the gradients of $\mathbf{u}$ \cite{Narozhny2019}. It will not be important for transport in a macroscopic sample \cite{Lucas2018}. The energy current density is proportional to the momentum
density $\mathbf{J}^{E}=v_{F}^{2}\mathbf{\pi}$
due to the linear dispersion relation of the WSM. Substituting the
above expressions in Eq. (\ref{eq:momentum conservation}), using
the energy conservation Eq. (\ref{eq:energy conservation}) and neglecting viscosity, we obtain
the Euler equation

\begin{eqnarray}&&
w\left(\frac{\partial}{\partial t}+\mathbf{u}\cdot\mathbf{\nabla}_{r}\right)\mathbf{u}+v_{F}^{2}\mathbf{\nabla}_{r}p+\mathbf{u}\frac{\partial p}{\partial t}-v_{F}^{2}en\mathbf{E} \nonumber \\&&
+e\left(\mathbf{E}\cdot\mathbf{J}^{n}\right)\mathbf{u}  \nonumber \\&&
=\sum_{l}\intop\left(d^{3}k\right)\mathbf{k}I_{\mathrm {e-imp}}\left[f_{l}(t,\mathbf{r},\mathbf{k})\right].\label{eq:Euler eq before integrating coll integral}
\end{eqnarray}

The RHS describes momentum relaxation by disordered impurities, and it is given
by

\begin{equation}
\sum_{l}\intop\left(d^{3}k\right)\mathbf{k}I_{\mathrm{e-imp}}\left[f_{l}(t,\mathbf{r},\mathbf{k})\right]=-\frac{\mathbf{u}}{\bar{\tau}_{\parallel}^{\textrm{el}}}\frac{w}{v_{F}^{2}},
\end{equation}
where the effective elastic transport time $\bar{\tau}_{\parallel}^{\textrm{el}}$
is calculated in the next section. Finally, using the thermodynamic
relation $dp=nd\mu+sdT$ we obtain the form of the Euler equation
written in the main text, Eq. (\ref{eq:Euler equation main}). For the
calculation of DC conductivities done in the main text, the time derivative
term is set to zero. We then find the boost velocity, to linear order
in the fields, to be given by

\begin{equation}
\mathbf{u}=\frac{v_{F}^{2}\bar{\tau}_{\textrm{\ensuremath{\parallel}}}^{\textrm{el}}}{w}\left[n\left(e\mathbf{E}+\mathbf{\nabla}_{r}\mu\right)-s\mathbf{\nabla}_{r}T\right].\label{eq:boost velocity linear response}
\end{equation}

We note that here, the boost velocity is limited only by the disorder.
In a different setup, the boost velocity may be limited by other scales, such as a finite frequency
of the fields. In this case one should replace $\bar{\tau}_{\parallel}^{\textrm{el}}\rightarrow\bar{\tau}_{\parallel}^{\textrm{el}}/\left(1-i\omega\bar{\tau}_{\parallel}^{\textrm{el}}\right)$
in the above equation. If the sample size is smaller than the elastic transport length, it should be used as a cut-off for  the momentum density zero-mode.

\section{Elastic transport scattering time\label{sec:Disorder-transport-scattering}}

The disorder collision integral is given by 
(omitting node and band indices
 as the scattering is elastic and we assumed
no internode scattering by the disorder \footnote{Note that electron-electron internode scattering exists
and is assumed to be much faster than the disorder scattering rate. Therefore, the assumption of no
internode scattering by the disorder is not crucial, and relaxing it
only modifies ${\tau}_{\parallel}^{\textrm{el}}$ by a numerical factor})

\begin{equation}
I_{\mathrm {e-imp}}\left[f_{\mathbf{k}}\right]=-\intop\left(d^{3}k'\right)\left(f_{\mathbf{k}}-f_{\mathbf{k}'}\right)w_{\mathbf{k} \mathbf{k}'},
\end{equation}

Where $w_{\mathbf{k} \mathbf{k}'}=2\pi\left|V_{\mathbf{k} \mathbf{k}'}\right|^{2}\delta(\epsilon_{\mathbf{k}}-\epsilon_{\mathbf{k}'})$
is the scattering rate, given by Fermi's golden rule. The elastic
transport time at a given energy is given by taking $f_{\mathbf{k}}\propto\hat{k}$, 

\begin{equation}
\frac{\hat{k}}{\tau_{\parallel}^{\textrm{el}}(\epsilon)}=-\intop\left(d^{3}k'\right)\left(\hat{k}-\hat{k}'\right)w_{\mathbf{k}\mathbf{k}'}=\frac{2\pi\gamma\nu(\epsilon)}{3}\hat{k},
\end{equation}

where $\nu(\epsilon)$ is the density of states of the single Weyl node at energy $\epsilon$
and $\gamma$ is the amplitude of the short-ranged disorder potential correlator
[Eq. (\ref{eq:Gaussian disorder})]. Note that due to the spinor structure,
low-angle scattering is favored (The matrix element $V_{\mathbf{k}\mathbf{k}'}$ contains the product of the Bloch eigenfunctions at $\mathbf{k},\mathbf{k}'$),
making the transport time larger than the elastic lifetime by a factor of $3/2$.

In the derivation of the Euler equation [Eq. (\ref{eq:Euler eq before integrating coll integral})],
the elastic transport rate $1/\tau_{\parallel}^{\textrm{el}}(\epsilon)$
is thermally averaged:

\begin{widetext}
\begin{equation}
\sum_{l}\intop\left(d^{3}k\right)\mathbf{k}I_{\mathrm {e-imp}}\left[f_{l}(t,\mathbf{r},\mathbf{k})\right]=-\mathbf{u}\sum_{l}\intop\left(d^{3}k\right)\left(-\frac{\partial f^{0}(\epsilon_{l \mathbf{k}})}{\partial\epsilon}\right)\frac{k^{2}}{3\tau_{\parallel}^{\textrm{el}}(\epsilon_{l\mathbf{k}})},
\end{equation}

where we linearized $f$ in the boost velocity, $f\approx f_{0}-\frac{\partial f_{0}}{\partial\epsilon}\mathbf{u}\cdot\mathbf{k}$.
We write the above expression by defining a thermally averaged transport
rate $1/\bar{\tau}_{\parallel}^{\textrm{el}}$:

\begin{equation}
\sum_{l}\intop\left(d^{3}k\right)\mathbf{k}I_{\mathrm {e-imp}}\left[f_{l}(t,\mathbf{r},\mathbf{k})\right]=-\frac{\mathbf{u}}{\bar{\tau}_{\parallel}^{\textrm{el}}}\sum_{l}\intop\left(d^{3}k\right)\left(-\frac{\partial f^{0}(\epsilon_{l\mathbf{k}})}{\partial\epsilon}\right)\frac{k^{2}}{3}=-\frac{\mathbf{u}}{\bar{\tau}_{\parallel}^{\textrm{el}}}\frac{w}{v_{F}^{2}},
\end{equation}

where the averaged transport rate can be seen to be given by

\begin{equation}
\frac{1}{\bar{\tau}_{\parallel}^{\textrm{el}}}=\frac{\sum_{l}\intop\left(d^{3}k\right)\left(-\frac{\partial f^{0}(\epsilon_{l\mathbf{k}})}{\partial\epsilon}\right)\frac{k^{2}}{\tau_{\parallel}^{\textrm{el}}(\epsilon_{l\mathbf{k}})}}{\sum_{l}\intop\left(d^{3}k\right)\left(-\frac{\partial f^{0}(\epsilon_{l\mathbf{k}})}{\partial\epsilon}\right)k^{2}}.
\end{equation}

Explicit calculation yields

\begin{equation}
\frac{1}{\bar{\tau}_{\parallel}^{\textrm{el}}}=\frac{1}{\tau_{\parallel}^{\textrm{el}}(\epsilon=T)}\frac{\left(\beta\mu\right)^{6}+5\pi^{2}\left(\beta\mu\right)^{4}+7\pi^{4}\left(\beta\mu\right)^{2}+\frac{31\pi^{6}}{21}}{\left(\beta\mu\right)^{4}+2\pi^{2}\left(\beta\mu\right)^{2}+\frac{7\pi^{4}}{15}}\approx\begin{cases}
\frac{155\pi^{2}}{49}1/\tau_{\parallel}^{\textrm{el}}(\epsilon=T) & \mu\ll T,\\
1/\tau_{\parallel}^{\textrm{el}}(\epsilon=\mu) & \mu\gg T.
\end{cases}
\end{equation}

Interestingly, for low temperatures $\left(\bar{\tau}_{\parallel}^{\textrm{el}}\right)^{-1}$
is greatly enhanced (factor of $\sim30$) compared to the naive estimation
$\left(\bar{\tau}_{\parallel}^{\textrm{el}}\right)^{-1}\approx\left(\tau_{\parallel}^{\textrm{el}}(\epsilon=T)\right)^{-1}$;
the high-energy electrons, which scatter faster, are given a higher
weight in the momentum relaxation integral.

\section{Experimental realizations of electron hydrodynamics in semimetals\label{sec:Experimental}}

Here we review some experimental properties of semimetals exhibiting (or being close to) the hydrodynamic regime.
In particular, we discuss the scales of the different scattering mechanisms
and their compatibility with our simple model, which neglects electron-phonon  scattering and 
assumes electron-electron scattering time to be the shortest.

Signatures of the hydrodynamic regime have been reported in the type-II, time-reversal
symmetric WSM WP$_2$ \cite{Gooth2018, Jaoui2018},
and in the closely related but non-Weyl phase of WTe$_2$ \cite{Vool2021, Amit2022}.
We summarize  the values  of scattering
lengths  based on  recent  experiments and \textit{ab initio} calculations in 
Table \ref{table:experimental-lengths}.

It is worth mentioning that \textit{ab initio} calculations \cite{Coulter2018,Vool2021} 
for both  materials show that the bare Coulomb interaction is strongly screened by the free carriers. 
The effective electron-electron interaction emerges due to the exchange of virtual phonons.
We emphasize that this virtual phonon-mediated interaction strictly conserves the energy and momentum of the electronic system, as opposed to a real phonon absorption/emission process (the latter corresponds to $\ell^\textrm{e-ph}$ in Table \ref{table:experimental-lengths}).
Since the microscopic origins of the electron-electron interaction are not consequential
for our analysis, this mechanism still leads to hydrodynamics of the same type as we consider.
The only influence of the microscopic origin of the electron-electron interaction is incorporated in the 
dependence of $\tau^\textrm{e-e}$ on the temperature and chemical potential.
Such an analysis in the case of phonon-mediated electron-electron interaction has recently been done in Ref. \cite{Bernabeu2022}.

\begin{center}
	
	\begin{table}[h]
		\begin{tabular}{ c c c c c c }
			\hline \hline
			\textbf{Material} & \textbf{References} & \textbf{\textit{T} [K]} & $\mathbf{\ell^\textrm{e-e} [\mu m]}$ & $\mathbf{\ell^\textrm{e-ph} [\mu m]}$ & $\mathbf{\ell^\textrm{e-imp} [\mu m]}$ \\
			\hline
			\multirow{4}{*}{WTe$_2$} & \multirow{4}{120pt}{\centering{Experimental realizations in Refs. \cite{Vool2021, Amit2022},
					\textit{ab initio} calculations in Ref. \cite{Vool2021} [*]}}
				 & 5 & ~25 & ~150 & \multirow{4}{12pt}{\centering{1.8}} \\
			&& 10 & ~2.5 & ~7.5 & \\
			&& 15 & ~1 & ~2 & \\
			&& 50 & ~0.1 & ~0.1 & \\
			\hline
			\multirow{4}{*}{WP$_2$} & \multirow{4}{120pt}{\centering{Experimental realizations in Refs. \cite{Gooth2018, Jaoui2018},
					\textit{ab initio} calculations in Ref. \cite{Coulter2018}}} & 5 & ~200 & ~30 & \multirow{4}{12pt}{\centering{~100 \cite{Gooth2018}}} \\
			&& 10 & ~9 & ~5 & \\
			&& 15 & ~1.5 & ~1.5 & \\
			&& 50 & ~0.03 & ~0.04 & \\
			\hline
			\multirow{3}{*}{Sb} & \multirow{3}{130pt}{\centering{Experimental and fitted parameters from Ref. \cite{Jaoui2021}}} & 3.5 & ~1000 & \multirow{3}{100pt}{\centering{{[}**]}} & \multirow{3}{70pt}{\centering{~200 \linebreak (cleanest sample)}} \\
			&&&&& \\
			&& 7.5 & ~300 & & \\
			\hline \hline
		\end{tabular}
		\caption{Scattering lengths for different materials, collected from other works.
			Data from Ref. \cite{Vool2021} are \textit{ab initio} calculations taken from Fig. 4. Data from Ref. \cite{Coulter2018}
			are taken from Fig. 2, with $v_F=1.4 \cdot 10^5 m/s$ to convert from scattering times to lengths. For Sb,
			the lengths $\ell^\textrm{e-e}$ are the Principi-Vignale formula fitted values of Fig. 6 in Ref. \cite{Jaoui2021}.
			In both calculations for WP$_2$ and WTe$_2$,
			the dominant electron-electron scattering is due to the phonon-mediated interaction. \\
   		    {[}*] Calculations in Ref. \cite{Amit2022} suggest that $\ell^\textrm{e-e}$ in WTe$_2$ may be smaller due to enhanced Coulomb scattering. \\
   	    	{[}**] Ref. \cite{Jaoui2021} argues negligible e-ph in Sb at the measured temperatures. The argument is expecting a Bloch-Grüneisen contribution to the resistivity
   	    	of $\rho_\textrm{BG} \sim T^5$ from phonons, while the data shows a negligible $\sim T^5$ term compared to the constant and $\sim T^2$ parts of the resistivity.}
		
		\label{table:experimental-lengths}
	\end{table}
\end{center}

By analyzing the experimental data summarized in Table \ref{table:experimental-lengths} one concludes that  though the hydrodynamic regime ($l^{\rm{e-e}} <l^{\rm{imp}}<l^{{\rm e-ph}}$) can be achieved, 
the separation between the scattering scales is not large.
Therefore the majority of experiments up to date are on the verge of the formation of the hydrodynamic regime.
Both electron-impurity and electron-phonon scattering can mask  the hydrodynamic effects.

Among the effects that fit into the hydrodynamic picture are  the  experiments in WP$_2$ that have reported a low Lorentz ratio \cite{Gooth2018, Jaoui2018},
in agreement with the hydrodynamic result [Eq. (\ref{eq:Lorenz_xx_degenerate})] at the
degenerate regime ($\left| \mu/T \right| \gg 1$).
Similar behavior has been also shown in antimony (Sb) \cite{Jaoui2021}.
However, in the antimony experiment, the impurity scattering length was comparable with the
electron-electron scattering length.  Therefore, the comparison with formula Eq. (\ref{eq:Lorenz_xx_degenerate}) 
should be performed using its modified form, the Principi-Vignale result, replacing
$\tau^{\textrm{e-e}}/\bar{\tau}_{\parallel}^{\textrm{el}} \rightarrow
 \tau^{\textrm{e-e}} / \left( \tau^{\textrm{e-e}} + \bar{\tau}_{\parallel}^{\textrm{el}} \right)$
 \cite{Principi2015}.

It should be noted that a low Lorenz ratio can also result from other mechanisms.
Particularly, due to  different relaxation rates for the electric and energy currents \cite{Jaoui2018}.
Small-angle inelastic momentum-relaxing scattering (i.e. electron-phonon or Umklapp electron-electron)
relaxes the energy current much  faster than the electric current (in the degenerate regime,
where the electric current is nearly equivalent to the total momentum \cite{Principi2015}).

It is worth mentioning that electron-electron Umklapp scattering mechanism may be relevant for
WP$_2$. This is due to electron and hole Fermi pockets separated in momentum
space, enhancing electron-electron Umklapp scattering \cite{Jaoui2018}. In this case,
one has to distinguish between the normal (momentum-conserving)
and Umklapp (momentum-relaxing) electron-electron scattering rates. In our model, $\tau^{\textrm{e-e}}$
denotes only the normal part of the e-e collisions. Significant Umklapp scatterings would contribute to
the relaxation of the momentum density as well as the electric and heat currents.


It is also worth to mention that the dominant mechanism 
of phonon decay over a wide temperature range in WP$_2$ is via phonon-electron
scattering, rather than phonon-phonon scattering\cite{Osterhoudt2021}.
This implies that the electron and phonon degrees of freedom in this material are strongly coupled.
In this case, the  energy and momentum of the
coupled electron-phonon fluid are conserved \cite{Osterhoudt2021, Levchenko2020}.
This regime is outside the scope of our work.




\section{Thermodynamic quantities}

Here we present some results for thermodynamics quantities for an isolated
Weyl node with 3D linear spectrum $H=v_F \mathbf{\sigma} \cdot \mathbf{k}$.

The grand potential of the system is given by \cite{Landau1959}

\begin{align}
\Omega(T,V,\mu) & =-\frac{1}{\beta}\sum_{b}\prod_{\mathbf{k}}\left[1+\exp\left(\beta\left(\mu-\epsilon_{b\mathbf{k}}\right)\right)\right]=-\frac{1}{\beta}\sum_{b\mathbf{k}}\ln\left[1+\exp\left(\beta\left(\mu-\epsilon_{b\mathbf{k}}\right)\right)\right]\nonumber \\
 & =\frac{V}{\pi^{2}\beta^{4}v_{F}^{3}}\left[\mathrm{Li}_{4}(-e^{\beta\mu})+\mathrm{Li}_{4}(-e^{-\beta\mu})\right],
\end{align}

where
\begin{equation}
\mathrm{Li}_{s}(-z)=-\frac{1}{\left(s-1\right)!}\intop_{0}^{\infty}\frac{t^{s-1}}{z^{-1}e^{t}+1}dt
\end{equation}

is the polylogarithm function. From the grand potential one can derive
the thermodynamic quantities:

\begin{align}
p & =-\frac{\Omega}{V},\\
N & =-\frac{\partial\Omega}{\partial\mu}=-\frac{V}{\pi^{2}\beta^{3}v_{F}^{3}}\left[\mathrm{Li}_{3}(-e^{\beta\mu})-\mathrm{Li}_{3}(-e^{-\beta\mu})\right],\\
S & =-\frac{\partial\Omega}{\partial T}=\beta\left(4pV-\mu N\right).
\end{align}

In terms of densities we find

\begin{align}
w & =\frac{1}{6\pi^{2}\beta^{4}v_{F}^{3}}\left[\left(\beta\mu\right)^{4}+2\pi^{2}\left(\beta\mu\right)^{2}+\frac{7\pi^{4}}{15}\right],\\
s & =\frac{1}{6\beta^{3}v_{F}^{3}}\left[\left(\beta\mu\right)^{2}+\frac{7\pi^{2}}{15}\right],\\
n & =\frac{1}{6\pi^{2}\beta^{3}v_{F}^{3}}\beta\mu\left[\left(\beta\mu\right)^{2}+\pi^{2}\right].
\end{align}

In the system we consider, assuming $\left| \mu \right|,T \ll \left( \Delta_k v_F \right)$, we extend the momentum integration for each Weyl node to infinity and get that the extensive quantities are simply the above results multiplied by the number of nodes $N_W=2$. However, there is a subtlety  here  if  the boost velocity is finite. Since the boost velocity $\mathbf{u}$ of the whole system couples to the quasimomentum $\mathbf{k}$, one has to consistently choose the same origin in k-space for both nodes. This effectively leads to the two nodes equilibrating to different chemical potentials, $\mu_\eta = \mu + \eta \mathbf{u} \cdot \mathbf{\Delta}_k/2$. Therefore, an external field along the node separation axis causes a charge drift between the nodes. For the quantities we consider, the  effect can be disregarded on the  linear response level.

The density of states of a single Weyl node is given by

\begin{equation}
\nu(\epsilon)=\frac{\epsilon^{2}}{2\pi^{2}v_{F}^{3}}.
\end{equation}

In the above calculations, we used the following useful identities
for the polylogarithms functions \cite{Lewin1981}:

\begin{align}
\mathrm{Li}_{2}(-e^{x})+\mathrm{Li}_{2}(-e^{-x}) & =-\frac{x^{2}}{2}-\frac{\pi^{2}}{6},\\
\mathrm{Li}_{3}(-e^{x})-\mathrm{Li}_{3}(-e^{-x}) & =-\frac{x^{3}}{6}-\frac{\pi^{2}}{6}x,\\
\mathrm{Li}_{4}(-e^{x})+\mathrm{Li}_{4}(-e^{-x}) & =-\frac{1}{24}x^{4}-\frac{\pi^{2}}{12}x^{2}-\frac{7\pi^{4}}{360},\\
\mathrm{Li}_{6}(-e^{x})+\mathrm{Li}_{6}(-e^{-x}) & =-\frac{1}{6!}\left(x^{6}+5\pi^{2}x^{4}+7\pi^{4}x^{2}+\frac{31\pi^{6}}{21}\right).
\end{align}

\section{Calculation of dissipative linear response coefficients\label{sec:Calculation-of-dissipative}}

In this part, we calculate the dissipative response coefficients by
approximately solving the electron-electron collision integral. To
the leading order in $\bar{\tau}_{\parallel}^{\textrm{el}}/\tau^{\textrm{e-e}}$,
the disorder collision integral can be neglected, and from hereon
we take $I\left[f\right]\approx I_{\textrm{e-e}}\left[f\right]$.

We write the distribution function as $f=f^{\textrm{FD}}+\delta f$,
where $f^{\textrm{FD}}$ is a local equilibrium Fermi-Dirac distribution
and $\delta f$ is the non-equilibrium dissipative correction. Let
us parametrize this correction by

\begin{equation}
\delta f_{l}(\mathbf{k})=g(\epsilon_{l\mathbf{k}})h_{l}(\mathbf{k}),\label{eq:dist fun dissipative correction}
\end{equation}

where $h(\mathbf{k})$ is any function and

\begin{equation}
g(\epsilon)=\sqrt{-\frac{1}{\beta}\frac{\partial f^{0}}{\partial\epsilon}}=\frac{1}{2\cosh\left(\frac{\epsilon-\mu}{2T}\right)}.
\end{equation}

The linearized Boltzmann equation reads

\begin{equation}
-\beta g^{2}(\epsilon_{l\mathbf{k}})\left[\mathbf{v}_{l\mathbf{k}}\cdot\left(e\mathbf{E}-\mathbf{\nabla}\mu\right)-\frac{\mathbf{v}_{l\mathbf{k}}\left(\epsilon_{l\mathbf{k}}-\mu\right)}{T}\cdot\mathbf{\nabla}T\right]=I\left[\delta f_{l}(\mathbf{k})\right].
\end{equation}

In order to solve the Boltzmann equation, one has to invert the collision
integral. However, the e-e collision integral has zero-modes, and
so first we must project the Boltzmann equation to the subspace orthogonal
to these zero-modes. The zero-modes correspond to the conserved quantities
of the collision integral: charge, momentum and energy densities,
and are given by

\begin{equation}
\lambda_{l}^{(1)}(\mathbf{k})=1,\qquad\lambda_{l}^{(2)}(\mathbf{k})=\mathbf{k},\qquad\lambda_{l}^{(3)}(\mathbf{k})=\epsilon_{l\mathbf{k}}.\label{eq:zero modes of coll. integral}
\end{equation}

We linearize the collision integral, and define the inner product
space

\begin{equation}
\left\langle \psi\vert\phi\right\rangle \equiv\sum_{l}\intop\left(d\mathbf{k}\right)\psi_{l}(\mathbf{k})g^{2}(\epsilon_{l\mathbf{k}})\phi_{l}(\mathbf{k}).\label{eq:inner space product}
\end{equation}

We symmetrize the collision integral in this inner product space by
defining

\begin{equation}
\mathscr{\mathcal{I}}_{l\mathbf{k},l'\mathbf{k}'}\equiv\frac{1}{g(\epsilon_{l\mathbf{k}})}I_{l\mathbf{k},l'\mathbf{k}'}g(\epsilon_{l'\mathbf{k}'}).
\end{equation}

Substituting the symmetrized collision operator in the Boltzmann equation,
we can invert it in the subspace orthogonal to the zero-modes and
find

\begin{equation}
h_{l}(\mathbf{k})=-\beta\sum_{l'}\intop\left(d\mathbf{k}'\right)\mathscr{\mathcal{I}}_{l\mathbf{k},l'\mathbf{k}'}^{-1}g(\epsilon_{l'\mathbf{k'}})\left[\mathbf{v}_{l'\mathbf{k}'}\cdot\left(e\mathbf{E}-\mathbf{\nabla}\mu\right)-\frac{\mathbf{v}_{l'\mathbf{k}'}\left(\epsilon_{l'\mathbf{k}'}-\mu\right)}{T}\mathbf{\nabla}T\right].
\end{equation}

At this stage, we approximate the inverse e-e collision integral to
be simply given by $\mathscr{\mathcal{I}}_{l\mathbf{k},l'\mathbf{k}'}^{-1}=-\tau^{\textrm{e-e}}\delta\left(\mathbf{k}-\mathbf{k}'\right)\delta_{ll'}$.
We then find the dissipative corrections to the currents,

\begin{equation}
\mathbf{J}_{\textrm{diss.}}^{i}\equiv\sum_{l}\intop\left(d\mathbf{k}\right)\mathbf{j}_{l\mathbf{k}}^{i}\delta f_{l\mathbf{k}}=\beta\tau^{\textrm{e-e}}\left\langle \mathbf{j}^{i}\vert\mathbf{v}\cdot\left[e\mathbf{E}-\mathbf{\nabla}\mu-\frac{\left(\epsilon-\mu\right)}{T}\mathbf{\nabla}T\right]\right\rangle ,\label{eq:dissipative currents inner product}
\end{equation}
\end{widetext}

where $i=1,2$ indicate the electric and thermal currents, with their
operators given by

\begin{equation}
\mathbf{j}^{\textrm{1}}=e\mathbf{v}_{l\mathbf{k}},\qquad\qquad\qquad\mathbf{j}^{\textrm{2}}=\left(\epsilon_{l\mathbf{k}}-\mu\right)\mathbf{v}_{l\mathbf{k}}.
\end{equation}

In the RHS of Eq. (\ref{eq:dissipative currents inner product}), both
vectors are projected to the subspace orthogonal to the zero-modes.
Eq. (\ref{eq:dissipative currents inner product}) can be written more
compactly as

\begin{equation}
\label{dissipative_current1}
J_{\textrm{diss.,}\alpha}^{i}=L_{\textrm{diss.,}\alpha\beta}^{ij}F_{\beta}^{j},
\end{equation}

where $\mathbf{F}^{i}$ are generalized forces, $L_{\textrm{diss.}}^{ij}$
are the dissipative parts of the total thermoelectric response tensors
defined in the main text [Eqs. (\ref{eq:general Ls eq 1}), (\ref{eq:general Ls eq 2})],
and explicitly:

\begin{align}
\mathbf{F}^{\textrm{1}} & =\mathbf{E}-\frac{\mathbf{\nabla}\mu}{e},\qquad\qquad\qquad\mathbf{F}^{\textrm{2}}=-\mathbf{\nabla}T,\\
L_{\textrm{diss.,}\alpha\beta}^{ij} & =-\beta\left\langle j_{\alpha}^{i}\vert\mathscr{\hat{\mathcal{I}}}^{-1}j_{\beta}^{j}\right\rangle .\label{eq:dissipative L-s}
\end{align}

By computing Eq. (\ref{eq:dissipative L-s}) we find,

\begin{align}
\label{dissipative_current2}
L_{\textrm{diss.,}xx}^{\textrm{11}} & =\sigma_{xx}^{\textrm{\textrm{diss.}}}=e^{2}v_{F}^{2}\left(\frac{1}{3}\frac{\partial n}{\partial\mu}-\frac{n^{2}}{w}\right)\tau^{\textrm{e-e}},\\
L_{\textrm{diss.,}xx}^{\textrm{21}} & =TL_{\textrm{diss,}xx}^{\textrm{12}}=-\frac{\mu}{e}\sigma_{xx}^{\textrm{\textrm{diss.}}},\\
L_{\textrm{diss.,}xx}^{\textrm{22}} & =\frac{\mu^{2}}{e^{2}T}\sigma_{xx}^{\textrm{diss.}}.
\end{align}

In our approximation for the electron-electron collision integral,
there are no contributions to the off-diagonal dissipative response
coefficients, $L_{\textrm{diss.,}xy}^{ij}=0$. Interestingly, $\sigma_{xx}^{\textrm{diss.}}$
can be written by a Drude-like formula in the two opposite limits
of $\mu/T$:

\begin{equation}
\sigma_{xx}^{\textrm{diss.}}\approx\frac{e^{2}v_{F}^{2}}{3}\tau^{\textrm{e-e}}\times\begin{cases}
\frac{\partial n}{\partial\mu} & \left|\mu/T\right|\ll1,\\
\frac{T}{\mu^{2}}c_{v} & \left|\mu/T\right|\gg1,
\end{cases}
\end{equation}

with $c_{v}=T\partial s/\partial T$ being the heat capacity. This
result reflects that the momentum zero-mode is parallel to the thermal
current in the Dirac fluid regime, and to the particle current in
the Fermi liquid regime. Dissipation occurs via the orthogonal channels,
therefore $\sigma_{xx}^{\textrm{diss.}}$ is related to particle diffusion
in the Dirac fluid and heat diffusion in the Fermi liquid.

Lastly, we address the electron-electron scattering time $\tau^{\textrm{e-e}}$.
In Weyl semimetals, the electron-electron scattering rate can be estimated
by $1/\tau^{\textrm{e-e}}\sim N_{W}\alpha^{2}T\min\left(1,\frac{T}{\mu}\right)$
\cite{Burkov2011,Hosur2012}, where $N_{W}$ is the number of Weyl
nodes and $\alpha$ is the effective fine-structure constant, characterizing
the ratio between the typical Coulomb interaction energy scale and
the typical kinetic energy scale. The condition $\tau^{\textrm{e-e}}\ll\bar{\tau}_{\parallel}^{\textrm{el}}$
gives a condition for the maximal strength of the disorder,

\begin{equation}
\gamma\ll\frac{v_{F}^{3}\alpha^{2}T^{2}}{\max(\mu^{3},T^{3})}.
\end{equation}

\section{Seebeck coefficient for non-interacting electrons \label{sec:Seebeck-coefficient-for}}

Here, we calculate the temperature dependence of the Seebeck coefficient
in a non-interacting system with particle-hole symmetry. While the
calculation is standard for the Fermi-liquid regime ($\mu\gg T$)
\cite{ashcroft1976solid,Behnia2004,Strunk2021}, for the Dirac fluid
regime ($\mu\ll T$) it is less commonly found in the literature,
and so we carry it here for comparison with the hydrodynamic regime.
For simplicity, we focus on an isotropical system and omit the tensor
structure (in space) of the response coefficients. For non-interacting
electrons, the thermoelectric response $L^{\textrm{12}}$ can be calculated
from the Mott relation

\begin{equation}
L^{\textrm{12}}=\frac{1}{e}\intop d\epsilon\frac{\epsilon-\mu}{T}\sigma^{\textrm{}}(\epsilon)\left(-\frac{\partial f^{0}}{\partial\epsilon}\right),
\end{equation}

where

\begin{equation}
\sigma(\epsilon)=e^{2}\intop\left(d\mathbf{k}\right)\delta(\epsilon-\epsilon_{\mathbf{k}})v^{2}(\mathbf{k})\tau(\mathbf{k})
\end{equation}

is the electric conductivity at zero temperature. For an isotropic
system,

\begin{equation}
\sigma(\epsilon)=\frac{e^{2}v_{F}^{2}(\epsilon)\nu(\epsilon)}{d}\tau(\epsilon),
\end{equation}

where $d$ is the number of dimensions. One can utilize the Sommerfeld
expansion

\begin{eqnarray}&&
\intop_{-\infty}^{\infty}K(\epsilon)\left(-\frac{\partial f}{\partial\epsilon}\right)d\epsilon
\nonumber \\&&
=K(\mu)+\sum_{n=1}^{\infty}a_{n}\left(T\right)^{2n}\frac{d^{2n}}{d\epsilon^{2n}}K(\epsilon)\vert_{\epsilon=\mu},
\end{eqnarray}

where $a_{n}$ are dimensionless coefficients, the first ones given
by $a_{1}=\pi^{2}/6,$ $a_{2}=7\pi^{4}/360$. Writing $v_{F}^{2}(\epsilon)\nu(\epsilon)\sim\epsilon^{\alpha}$
and $\tau\sim\epsilon^{\beta}$ (for example, in 3D metals with scattering
due to short-ranged impurities, $\alpha=3/2,\beta=-1/2$) one finds
for the Fermi-liquid regime, to leading order in $\epsilon_{F}/T$,

\begin{align}
L^{11} & =\sigma(\epsilon_{F}),\\
L^{12} & =\frac{\pi^{2}}{3e}T\left.\frac{d\sigma}{d\epsilon}\right|_{\epsilon=\epsilon_{F}}=\frac{\pi^{2}}{3e}\frac{T}{\epsilon_{F}}\left(\alpha+\beta\right)\sigma(\epsilon_{F}),
\end{align}

leading to the Seebeck coefficient

\begin{equation}
S=\frac{L^{12}}{L^{11}}=\frac{\pi^{2}}{3e}\frac{T}{\epsilon_{F}}\left(\alpha+\beta\right).\label{eq:Seebeck coefficient Fermi liquid}
\end{equation}

The thermopower of WSMs in the linear dispersion regime was studied
by Lundgren \textit{et al}. \cite{Lundgren2014}. In WSMs, $\alpha=2$. Focusing on the
leading order in $\epsilon_{F}/T$, their results imply $\beta=2$
for scattering due to charged impurities and $\beta=-2$ for short-ranged
impurities. Plugging these values in Eq. (\ref{eq:Seebeck coefficient Fermi liquid})
gives agreement with Ref. \cite{Lundgren2014}.

In the Dirac fluid regime, one can still perform the Sommerfeld expansion
as long as $\sigma(\epsilon)$ is a polynomial function of $\epsilon$.
Assuming particle-hole symmetry, $\sigma(\epsilon)$ is even in energy.
Denoting the highest power of $\epsilon$ in the expansion of $\sigma(\epsilon)$
as $2\gamma$, one finds, to the leading order in $T/\mu$,

\begin{align}
L^{11} & =a_{\gamma}\left(2\gamma\right)!\left(\frac{T}{\mu}\right)^{2\gamma}\sigma(\mu),\\
L^{12} & =\frac{1}{e}a_{\gamma}\left(2\gamma\right)!\left(\frac{T}{\mu}\right)^{2\gamma-1}\sigma(\mu),
\end{align}


leading to

\begin{equation}
S=\frac{1}{e}\frac{\mu}{T}.
\end{equation}

We see that the Seebeck coefficient in the non-interacting Dirac fluid
regime scales as $\mu/T$, in contrast to the
hydrodynamic regime where it scales as $s/n\simeq T/\mu$. 

\twocolumngrid

\bibliography{citations}

\begin{thebibliography}{64}%
\makeatletter
\providecommand \@ifxundefined [1]{%
 \@ifx{#1\undefined}
}%
\providecommand \@ifnum [1]{%
 \ifnum #1\expandafter \@firstoftwo
 \else \expandafter \@secondoftwo
 \fi
}%
\providecommand \@ifx [1]{%
 \ifx #1\expandafter \@firstoftwo
 \else \expandafter \@secondoftwo
 \fi
}%
\providecommand \natexlab [1]{#1}%
\providecommand \enquote  [1]{``#1''}%
\providecommand \bibnamefont  [1]{#1}%
\providecommand \bibfnamefont [1]{#1}%
\providecommand \citenamefont [1]{#1}%
\providecommand \href@noop [0]{\@secondoftwo}%
\providecommand \href [0]{\begingroup \@sanitize@url \@href}%
\providecommand \@href[1]{\@@startlink{#1}\@@href}%
\providecommand \@@href[1]{\endgroup#1\@@endlink}%
\providecommand \@sanitize@url [0]{\catcode `\\12\catcode `\$12\catcode
  `\&12\catcode `\#12\catcode `\^12\catcode `\_12\catcode `\%12\relax}%
\providecommand \@@startlink[1]{}%
\providecommand \@@endlink[0]{}%
\providecommand \url  [0]{\begingroup\@sanitize@url \@url }%
\providecommand \@url [1]{\endgroup\@href {#1}{\urlprefix }}%
\providecommand \urlprefix  [0]{URL }%
\providecommand \Eprint [0]{\href }%
\providecommand \doibase [0]{https://doi.org/}%
\providecommand \selectlanguage [0]{\@gobble}%
\providecommand \bibinfo  [0]{\@secondoftwo}%
\providecommand \bibfield  [0]{\@secondoftwo}%
\providecommand \translation [1]{[#1]}%
\providecommand \BibitemOpen [0]{}%
\providecommand \bibitemStop [0]{}%
\providecommand \bibitemNoStop [0]{.\EOS\space}%
\providecommand \EOS [0]{\spacefactor3000\relax}%
\providecommand \BibitemShut  [1]{\csname bibitem#1\endcsname}%
\let\auto@bib@innerbib\@empty
\bibitem [{\citenamefont {Levitov}\ and\ \citenamefont
  {Falkovich}(2016)}]{Levitov2016}%
  \BibitemOpen
  \bibfield  {author} {\bibinfo {author} {\bibfnamefont {L.}~\bibnamefont
  {Levitov}}\ and\ \bibinfo {author} {\bibfnamefont {G.}~\bibnamefont
  {Falkovich}},\ }\bibfield  {title} {\bibinfo {title} {Electron viscosity,
  current vortices and negative nonlocal resistance in graphene},\ }\href@noop
  {} {\bibfield  {journal} {\bibinfo  {journal} {Nat. Phys.}\ }\textbf
  {\bibinfo {volume} {12}},\ \bibinfo {pages} {672} (\bibinfo {year}
  {2016})}\BibitemShut {NoStop}%
\bibitem [{\citenamefont {Gurzhi}(1963)}]{Gurzhi1963}%
  \BibitemOpen
  \bibfield  {author} {\bibinfo {author} {\bibfnamefont {R.~N.}\ \bibnamefont
  {Gurzhi}},\ }\bibfield  {title} {\bibinfo {title} {Minimum of resistance in
  impurity free conductors},\ }\href@noop {} {\bibfield  {journal} {\bibinfo
  {journal} {Sov. Phys. JETP}\ }\textbf {\bibinfo {volume} {17}},\ \bibinfo
  {pages} {521} (\bibinfo {year} {1963})}\BibitemShut {NoStop}%
\bibitem [{\citenamefont {Ghahari}\ \emph {et~al.}(2016)\citenamefont
  {Ghahari}, \citenamefont {Xie}, \citenamefont {Taniguchi}, \citenamefont
  {Watanabe}, \citenamefont {Foster},\ and\ \citenamefont {Kim}}]{Ghahari2016}%
  \BibitemOpen
  \bibfield  {author} {\bibinfo {author} {\bibfnamefont {F.}~\bibnamefont
  {Ghahari}}, \bibinfo {author} {\bibfnamefont {H.~Y.}\ \bibnamefont {Xie}},
  \bibinfo {author} {\bibfnamefont {T.}~\bibnamefont {Taniguchi}}, \bibinfo
  {author} {\bibfnamefont {K.}~\bibnamefont {Watanabe}}, \bibinfo {author}
  {\bibfnamefont {M.~S.}\ \bibnamefont {Foster}},\ and\ \bibinfo {author}
  {\bibfnamefont {P.}~\bibnamefont {Kim}},\ }\bibfield  {title} {\bibinfo
  {title} {Enhanced thermoelectric power in graphene: Violation of the {Mott}
  relation by inelastic scattering},\ }\href@noop {} {\bibfield  {journal}
  {\bibinfo  {journal} {Phys. Rev. Lett.}\ }\textbf {\bibinfo {volume} {116}},\
  \bibinfo {pages} {136802} (\bibinfo {year} {2016})}\BibitemShut {NoStop}%
\bibitem [{\citenamefont {Moll}\ \emph {et~al.}(2016)\citenamefont {Moll},
  \citenamefont {Kushwaha}, \citenamefont {Nandi}, \citenamefont {Schmidt},\
  and\ \citenamefont {Mackenzie}}]{Moll2016}%
  \BibitemOpen
  \bibfield  {author} {\bibinfo {author} {\bibfnamefont {P.~J.~W.}\
  \bibnamefont {Moll}}, \bibinfo {author} {\bibfnamefont {P.}~\bibnamefont
  {Kushwaha}}, \bibinfo {author} {\bibfnamefont {N.}~\bibnamefont {Nandi}},
  \bibinfo {author} {\bibfnamefont {B.}~\bibnamefont {Schmidt}},\ and\ \bibinfo
  {author} {\bibfnamefont {A.~P.}\ \bibnamefont {Mackenzie}},\ }\bibfield
  {title} {\bibinfo {title} {Evidence for hydrodynamic electron flow in
  {PdCoO}\textsubscript{2}},\ }\href@noop {} {\bibfield  {journal} {\bibinfo
  {journal} {Science}\ }\textbf {\bibinfo {volume} {351}},\ \bibinfo {pages}
  {1061} (\bibinfo {year} {2016})}\BibitemShut {NoStop}%
\bibitem [{\citenamefont {Crossno}\ \emph {et~al.}(2016)\citenamefont
  {Crossno}, \citenamefont {Shi}, \citenamefont {Wang}, \citenamefont {Liu},
  \citenamefont {Harzheim}, \citenamefont {Lucas}, \citenamefont {Sachdev},
  \citenamefont {Kim}, \citenamefont {Taniguchi}, \citenamefont {Watanabe},
  \citenamefont {Ohki},\ and\ \citenamefont {Fong}}]{Crossno2016}%
  \BibitemOpen
  \bibfield  {author} {\bibinfo {author} {\bibfnamefont {J.}~\bibnamefont
  {Crossno}}, \bibinfo {author} {\bibfnamefont {J.~K.}\ \bibnamefont {Shi}},
  \bibinfo {author} {\bibfnamefont {K.}~\bibnamefont {Wang}}, \bibinfo {author}
  {\bibfnamefont {X.}~\bibnamefont {Liu}}, \bibinfo {author} {\bibfnamefont
  {A.}~\bibnamefont {Harzheim}}, \bibinfo {author} {\bibfnamefont
  {A.}~\bibnamefont {Lucas}}, \bibinfo {author} {\bibfnamefont
  {S.}~\bibnamefont {Sachdev}}, \bibinfo {author} {\bibfnamefont
  {P.}~\bibnamefont {Kim}}, \bibinfo {author} {\bibfnamefont {T.}~\bibnamefont
  {Taniguchi}}, \bibinfo {author} {\bibfnamefont {K.}~\bibnamefont {Watanabe}},
  \bibinfo {author} {\bibfnamefont {T.~A.}\ \bibnamefont {Ohki}},\ and\
  \bibinfo {author} {\bibfnamefont {K.~C.}\ \bibnamefont {Fong}},\ }\bibfield
  {title} {\bibinfo {title} {Observation of the {Dirac} fluid and the breakdown
  of the {Wiedemann-Franz} law in graphene},\ }\href@noop {} {\bibfield
  {journal} {\bibinfo  {journal} {Science}\ }\textbf {\bibinfo {volume}
  {351}},\ \bibinfo {pages} {1058} (\bibinfo {year} {2016})}\BibitemShut
  {NoStop}%
\bibitem [{\citenamefont {Gooth}\ \emph {et~al.}(2018)\citenamefont {Gooth},
  \citenamefont {Menges}, \citenamefont {Kumar}, \citenamefont {S{\"u}{\ss}},
  \citenamefont {Shekhar}, \citenamefont {Sun}, \citenamefont {Drechsler},
  \citenamefont {Zierold}, \citenamefont {Felser},\ and\ \citenamefont
  {Gotsmann}}]{Gooth2018}%
  \BibitemOpen
  \bibfield  {author} {\bibinfo {author} {\bibfnamefont {J.}~\bibnamefont
  {Gooth}}, \bibinfo {author} {\bibfnamefont {F.}~\bibnamefont {Menges}},
  \bibinfo {author} {\bibfnamefont {N.}~\bibnamefont {Kumar}}, \bibinfo
  {author} {\bibfnamefont {V.}~\bibnamefont {S{\"u}{\ss}}}, \bibinfo {author}
  {\bibfnamefont {C.}~\bibnamefont {Shekhar}}, \bibinfo {author} {\bibfnamefont
  {Y.}~\bibnamefont {Sun}}, \bibinfo {author} {\bibfnamefont {U.}~\bibnamefont
  {Drechsler}}, \bibinfo {author} {\bibfnamefont {R.}~\bibnamefont {Zierold}},
  \bibinfo {author} {\bibfnamefont {C.}~\bibnamefont {Felser}},\ and\ \bibinfo
  {author} {\bibfnamefont {B.}~\bibnamefont {Gotsmann}},\ }\bibfield  {title}
  {\bibinfo {title} {Thermal and electrical signatures of a hydrodynamic
  electron fluid in tungsten diphosphide},\ }\href
  {https://doi.org/10.1038/s41467-018-06688-y} {\bibfield  {journal} {\bibinfo
  {journal} {Nat. Commun.}\ }\textbf {\bibinfo {volume} {9}},\ \bibinfo {pages}
  {1} (\bibinfo {year} {2018})}\BibitemShut {NoStop}%
\bibitem [{\citenamefont {Jaoui}\ \emph {et~al.}(2018)\citenamefont {Jaoui},
  \citenamefont {Fauqu{\'e}}, \citenamefont {Rischau}, \citenamefont {Subedi},
  \citenamefont {Fu}, \citenamefont {Gooth}, \citenamefont {Kumar},
  \citenamefont {S{\"u}{\ss}}, \citenamefont {Maslov}, \citenamefont {Felser},\
  and\ \citenamefont {Behnia}}]{Jaoui2018}%
  \BibitemOpen
  \bibfield  {author} {\bibinfo {author} {\bibfnamefont {A.}~\bibnamefont
  {Jaoui}}, \bibinfo {author} {\bibfnamefont {B.}~\bibnamefont {Fauqu{\'e}}},
  \bibinfo {author} {\bibfnamefont {C.~W.}\ \bibnamefont {Rischau}}, \bibinfo
  {author} {\bibfnamefont {A.}~\bibnamefont {Subedi}}, \bibinfo {author}
  {\bibfnamefont {C.}~\bibnamefont {Fu}}, \bibinfo {author} {\bibfnamefont
  {J.}~\bibnamefont {Gooth}}, \bibinfo {author} {\bibfnamefont
  {N.}~\bibnamefont {Kumar}}, \bibinfo {author} {\bibfnamefont
  {V.}~\bibnamefont {S{\"u}{\ss}}}, \bibinfo {author} {\bibfnamefont {D.~L.}\
  \bibnamefont {Maslov}}, \bibinfo {author} {\bibfnamefont {C.}~\bibnamefont
  {Felser}},\ and\ \bibinfo {author} {\bibfnamefont {K.}~\bibnamefont
  {Behnia}},\ }\bibfield  {title} {\bibinfo {title} {Departure from the
  {Wiedemann-Franz} law in {WP$_2$} driven by mismatch in {T-square}
  resistivity prefactors},\ }\href@noop {} {\bibfield  {journal} {\bibinfo
  {journal} {npj Quantum Mater.}\ }\textbf {\bibinfo {volume} {3}},\ \bibinfo
  {pages} {64} (\bibinfo {year} {2018})}\BibitemShut {NoStop}%
\bibitem [{\citenamefont {Vool}\ \emph {et~al.}(2021)\citenamefont {Vool},
  \citenamefont {Hamo}, \citenamefont {Varnavides}, \citenamefont {Wang},
  \citenamefont {Zhou}, \citenamefont {Kumar}, \citenamefont {Dovzhenko},
  \citenamefont {Qiu}, \citenamefont {Garcia}, \citenamefont {Pierce},
  \citenamefont {Gooth}, \citenamefont {Anikeeva}, \citenamefont {Felser},
  \citenamefont {Narang},\ and\ \citenamefont {Yacoby}}]{Vool2021}%
  \BibitemOpen
  \bibfield  {author} {\bibinfo {author} {\bibfnamefont {U.}~\bibnamefont
  {Vool}}, \bibinfo {author} {\bibfnamefont {A.}~\bibnamefont {Hamo}}, \bibinfo
  {author} {\bibfnamefont {G.}~\bibnamefont {Varnavides}}, \bibinfo {author}
  {\bibfnamefont {Y.}~\bibnamefont {Wang}}, \bibinfo {author} {\bibfnamefont
  {T.~X.}\ \bibnamefont {Zhou}}, \bibinfo {author} {\bibfnamefont
  {N.}~\bibnamefont {Kumar}}, \bibinfo {author} {\bibfnamefont
  {Y.}~\bibnamefont {Dovzhenko}}, \bibinfo {author} {\bibfnamefont
  {Z.}~\bibnamefont {Qiu}}, \bibinfo {author} {\bibfnamefont {C.~A.}\
  \bibnamefont {Garcia}}, \bibinfo {author} {\bibfnamefont {A.~T.}\
  \bibnamefont {Pierce}}, \bibinfo {author} {\bibfnamefont {J.}~\bibnamefont
  {Gooth}}, \bibinfo {author} {\bibfnamefont {P.}~\bibnamefont {Anikeeva}},
  \bibinfo {author} {\bibfnamefont {C.}~\bibnamefont {Felser}}, \bibinfo
  {author} {\bibfnamefont {P.}~\bibnamefont {Narang}},\ and\ \bibinfo {author}
  {\bibfnamefont {A.}~\bibnamefont {Yacoby}},\ }\bibfield  {title} {\bibinfo
  {title} {Imaging phonon-mediated hydrodynamic flow in {WTe$_2$}},\
  }\href@noop {} {\bibfield  {journal} {\bibinfo  {journal} {Nat. Phys.}\
  }\textbf {\bibinfo {volume} {17}},\ \bibinfo {pages} {1216} (\bibinfo {year}
  {2021})}\BibitemShut {NoStop}%
\bibitem [{\citenamefont {Jaoui}\ \emph {et~al.}(2021)\citenamefont {Jaoui},
  \citenamefont {Fauqu{\'e}},\ and\ \citenamefont {Behnia}}]{Jaoui2021}%
  \BibitemOpen
  \bibfield  {author} {\bibinfo {author} {\bibfnamefont {A.}~\bibnamefont
  {Jaoui}}, \bibinfo {author} {\bibfnamefont {B.}~\bibnamefont {Fauqu{\'e}}},\
  and\ \bibinfo {author} {\bibfnamefont {K.}~\bibnamefont {Behnia}},\
  }\bibfield  {title} {\bibinfo {title} {Thermal resistivity and hydrodynamics
  of the degenerate electron fluid in antimony},\ }\href@noop {} {\bibfield
  {journal} {\bibinfo  {journal} {Nat. Commun.}\ }\textbf {\bibinfo {volume}
  {12}},\ \bibinfo {pages} {195} (\bibinfo {year} {2021})}\BibitemShut
  {NoStop}%
\bibitem [{\citenamefont {Aharon-Steinberg}\ \emph {et~al.}(2022)\citenamefont
  {Aharon-Steinberg}, \citenamefont {V{\"o}lkl}, \citenamefont {Kaplan},
  \citenamefont {Pariari}, \citenamefont {Roy}, \citenamefont {Holder},
  \citenamefont {Wolf}, \citenamefont {Meltzer}, \citenamefont {Myasoedov},
  \citenamefont {Huber}, \citenamefont {Yan}, \citenamefont {Falkovich},
  \citenamefont {Levitov}, \citenamefont {H{\"u}cker},\ and\ \citenamefont
  {Zeldov}}]{Amit2022}%
  \BibitemOpen
  \bibfield  {author} {\bibinfo {author} {\bibfnamefont {A.}~\bibnamefont
  {Aharon-Steinberg}}, \bibinfo {author} {\bibfnamefont {T.}~\bibnamefont
  {V{\"o}lkl}}, \bibinfo {author} {\bibfnamefont {A.}~\bibnamefont {Kaplan}},
  \bibinfo {author} {\bibfnamefont {A.~K.}\ \bibnamefont {Pariari}}, \bibinfo
  {author} {\bibfnamefont {I.}~\bibnamefont {Roy}}, \bibinfo {author}
  {\bibfnamefont {T.}~\bibnamefont {Holder}}, \bibinfo {author} {\bibfnamefont
  {Y.}~\bibnamefont {Wolf}}, \bibinfo {author} {\bibfnamefont {A.~Y.}\
  \bibnamefont {Meltzer}}, \bibinfo {author} {\bibfnamefont {Y.}~\bibnamefont
  {Myasoedov}}, \bibinfo {author} {\bibfnamefont {M.~E.}\ \bibnamefont
  {Huber}}, \bibinfo {author} {\bibfnamefont {B.}~\bibnamefont {Yan}}, \bibinfo
  {author} {\bibfnamefont {G.}~\bibnamefont {Falkovich}}, \bibinfo {author}
  {\bibfnamefont {L.~S.}\ \bibnamefont {Levitov}}, \bibinfo {author}
  {\bibfnamefont {M.}~\bibnamefont {H{\"u}cker}},\ and\ \bibinfo {author}
  {\bibfnamefont {E.}~\bibnamefont {Zeldov}},\ }\bibfield  {title} {\bibinfo
  {title} {Direct observation of vortices in an electron fluid},\ }\href@noop
  {} {\bibfield  {journal} {\bibinfo  {journal} {Nature}\ }\textbf {\bibinfo
  {volume} {607}},\ \bibinfo {pages} {74} (\bibinfo {year} {2022})}\BibitemShut
  {NoStop}%
\bibitem [{\citenamefont {Ashcroft}\ and\ \citenamefont
  {Mermin}(1976)}]{ashcroft1976solid}%
  \BibitemOpen
  \bibfield  {author} {\bibinfo {author} {\bibfnamefont {N.~W.}\ \bibnamefont
  {Ashcroft}}\ and\ \bibinfo {author} {\bibfnamefont {N.~D.}\ \bibnamefont
  {Mermin}},\ }\href@noop {} {\emph {\bibinfo {title} {Solid State Physics}}}\
  (\bibinfo  {publisher} {Holt, Rinehart and Winston},\ \bibinfo {year}
  {1976})\BibitemShut {NoStop}%
\bibitem [{\citenamefont {Fu}\ \emph {et~al.}(2020)\citenamefont {Fu},
  \citenamefont {Sun},\ and\ \citenamefont {Felser}}]{Fu2020}%
  \BibitemOpen
  \bibfield  {author} {\bibinfo {author} {\bibfnamefont {C.}~\bibnamefont
  {Fu}}, \bibinfo {author} {\bibfnamefont {Y.}~\bibnamefont {Sun}},\ and\
  \bibinfo {author} {\bibfnamefont {C.}~\bibnamefont {Felser}},\ }\bibfield
  {title} {\bibinfo {title} {Topological thermoelectrics},\ }\href@noop {}
  {\bibfield  {journal} {\bibinfo  {journal} {APL Mater.}\ }\textbf {\bibinfo
  {volume} {8}},\ \bibinfo {pages} {040913} (\bibinfo {year}
  {2020})}\BibitemShut {NoStop}%
\bibitem [{\citenamefont {Andersen}\ \emph {et~al.}(2019)\citenamefont
  {Andersen}, \citenamefont {Smith},\ and\ \citenamefont
  {Principi}}]{Andersen2019}%
  \BibitemOpen
  \bibfield  {author} {\bibinfo {author} {\bibfnamefont {T.~I.}\ \bibnamefont
  {Andersen}}, \bibinfo {author} {\bibfnamefont {T.~B.}\ \bibnamefont
  {Smith}},\ and\ \bibinfo {author} {\bibfnamefont {A.}~\bibnamefont
  {Principi}},\ }\bibfield  {title} {\bibinfo {title} {Enhanced photoenergy
  harvesting and extreme {Thomson} effect in hydrodynamic electronic systems},\
  }\href@noop {} {\bibfield  {journal} {\bibinfo  {journal} {Phys. Rev. Lett.}\
  }\textbf {\bibinfo {volume} {122}},\ \bibinfo {pages} {166802} (\bibinfo
  {year} {2019})}\BibitemShut {NoStop}%
\bibitem [{\citenamefont {Lucas}\ and\ \citenamefont {Fong}(2018)}]{Lucas2018}%
  \BibitemOpen
  \bibfield  {author} {\bibinfo {author} {\bibfnamefont {A.}~\bibnamefont
  {Lucas}}\ and\ \bibinfo {author} {\bibfnamefont {K.~C.}\ \bibnamefont
  {Fong}},\ }\bibfield  {title} {\bibinfo {title} {Hydrodynamics of electrons
  in graphene},\ }\href {https://doi.org/10.1088/1361-648X/AAA274} {\bibfield
  {journal} {\bibinfo  {journal} {J. Phys. Condens. Matter}\ }\textbf {\bibinfo
  {volume} {30}},\ \bibinfo {pages} {053001} (\bibinfo {year}
  {2018})}\BibitemShut {NoStop}%
\bibitem [{\citenamefont {Yan}\ and\ \citenamefont {Felser}(2017)}]{Yan2017}%
  \BibitemOpen
  \bibfield  {author} {\bibinfo {author} {\bibfnamefont {B.}~\bibnamefont
  {Yan}}\ and\ \bibinfo {author} {\bibfnamefont {C.}~\bibnamefont {Felser}},\
  }\bibfield  {title} {\bibinfo {title} {Topological materials: {Weyl}
  semimetals},\ }\href@noop {} {\bibfield  {journal} {\bibinfo  {journal}
  {Annu. Rev. Condens. Matter Phys.}\ }\textbf {\bibinfo {volume} {8}},\
  \bibinfo {pages} {337} (\bibinfo {year} {2017})}\BibitemShut {NoStop}%
\bibitem [{\citenamefont {Armitage}\ \emph {et~al.}(2018)\citenamefont
  {Armitage}, \citenamefont {Mele},\ and\ \citenamefont
  {Vishwanath}}]{Armitage2018}%
  \BibitemOpen
  \bibfield  {author} {\bibinfo {author} {\bibfnamefont {N.~P.}\ \bibnamefont
  {Armitage}}, \bibinfo {author} {\bibfnamefont {E.~J.}\ \bibnamefont {Mele}},\
  and\ \bibinfo {author} {\bibfnamefont {A.}~\bibnamefont {Vishwanath}},\
  }\bibfield  {title} {\bibinfo {title} {Weyl and {Dirac} semimetals in
  three-dimensional solids},\ }\href@noop {} {\bibfield  {journal} {\bibinfo
  {journal} {RMP}\ }\textbf {\bibinfo {volume} {90}},\ \bibinfo {pages}
  {015001} (\bibinfo {year} {2018})}\BibitemShut {NoStop}%
\bibitem [{\citenamefont {Xiao}\ \emph {et~al.}(2010)\citenamefont {Xiao},
  \citenamefont {Chang},\ and\ \citenamefont {Niu}}]{Xiao2010}%
  \BibitemOpen
  \bibfield  {author} {\bibinfo {author} {\bibfnamefont {D.}~\bibnamefont
  {Xiao}}, \bibinfo {author} {\bibfnamefont {M.~C.}\ \bibnamefont {Chang}},\
  and\ \bibinfo {author} {\bibfnamefont {Q.}~\bibnamefont {Niu}},\ }\bibfield
  {title} {\bibinfo {title} {Berry phase effects on electronic properties},\
  }\href {https://doi.org/10.1103/RevModPhys.82.1959} {\bibfield  {journal}
  {\bibinfo  {journal} {RMP}\ }\textbf {\bibinfo {volume} {82}},\ \bibinfo
  {pages} {1959} (\bibinfo {year} {2010})}\BibitemShut {NoStop}%
\bibitem [{\citenamefont {Burkov}\ and\ \citenamefont
  {Balents}(2011)}]{Burkov2011b}%
  \BibitemOpen
  \bibfield  {author} {\bibinfo {author} {\bibfnamefont {A.~A.}\ \bibnamefont
  {Burkov}}\ and\ \bibinfo {author} {\bibfnamefont {L.}~\bibnamefont
  {Balents}},\ }\bibfield  {title} {\bibinfo {title} {Weyl semimetal in a
  topological insulator multilayer},\ }\href
  {https://doi.org/10.1103/PhysRevLett.107.127205} {\bibfield  {journal}
  {\bibinfo  {journal} {Phys. Rev. Lett.}\ }\textbf {\bibinfo {volume} {107}},\
  \bibinfo {pages} {127205} (\bibinfo {year} {2011})}\BibitemShut {NoStop}%
\bibitem [{\citenamefont {Qin}\ \emph {et~al.}(2011)\citenamefont {Qin},
  \citenamefont {Niu},\ and\ \citenamefont {Shi}}]{Qin2011}%
  \BibitemOpen
  \bibfield  {author} {\bibinfo {author} {\bibfnamefont {T.}~\bibnamefont
  {Qin}}, \bibinfo {author} {\bibfnamefont {Q.}~\bibnamefont {Niu}},\ and\
  \bibinfo {author} {\bibfnamefont {J.}~\bibnamefont {Shi}},\ }\bibfield
  {title} {\bibinfo {title} {Energy magnetization and the thermal {Hall}
  effect},\ }\href
  {https://journals.aps.org/prl/abstract/10.1103/PhysRevLett.107.236601}
  {\bibfield  {journal} {\bibinfo  {journal} {Phys. Rev. Lett.}\ }\textbf
  {\bibinfo {volume} {107}},\ \bibinfo {pages} {236601} (\bibinfo {year}
  {2011})}\BibitemShut {NoStop}%
\bibitem [{\citenamefont {Haldane}(2004)}]{Haldane2004}%
  \BibitemOpen
  \bibfield  {author} {\bibinfo {author} {\bibfnamefont {F.~D.~M.}\
  \bibnamefont {Haldane}},\ }\bibfield  {title} {\bibinfo {title} {Berry
  curvature on the fermi surface: Anomalous hall effect as a topological
  fermi-liquid property},\ }\href
  {https://doi.org/10.1103/PhysRevLett.93.206602} {\bibfield  {journal}
  {\bibinfo  {journal} {Phys. Rev. Lett.}\ }\textbf {\bibinfo {volume} {93}},\
  \bibinfo {pages} {206602} (\bibinfo {year} {2004})}\BibitemShut {NoStop}%
\bibitem [{\citenamefont {Xiao}\ \emph {et~al.}(2006)\citenamefont {Xiao},
  \citenamefont {Yao}, \citenamefont {Fang},\ and\ \citenamefont
  {Niu}}]{Xiao2006}%
  \BibitemOpen
  \bibfield  {author} {\bibinfo {author} {\bibfnamefont {D.}~\bibnamefont
  {Xiao}}, \bibinfo {author} {\bibfnamefont {Y.}~\bibnamefont {Yao}}, \bibinfo
  {author} {\bibfnamefont {Z.}~\bibnamefont {Fang}},\ and\ \bibinfo {author}
  {\bibfnamefont {Q.}~\bibnamefont {Niu}},\ }\bibfield  {title} {\bibinfo
  {title} {Berry-phase effect in anomalous thermoelectric transport},\
  }\href@noop {} {\bibfield  {journal} {\bibinfo  {journal} {Phys. Rev. Lett.}\
  }\textbf {\bibinfo {volume} {97}},\ \bibinfo {pages} {026603} (\bibinfo
  {year} {2006})}\BibitemShut {NoStop}%
\bibitem [{\citenamefont {Gorbar}\ \emph {et~al.}(2017)\citenamefont {Gorbar},
  \citenamefont {Miransky}, \citenamefont {Shovkovy},\ and\ \citenamefont
  {Sukhachov}}]{Gorbar2017}%
  \BibitemOpen
  \bibfield  {author} {\bibinfo {author} {\bibfnamefont {E.~V.}\ \bibnamefont
  {Gorbar}}, \bibinfo {author} {\bibfnamefont {V.~A.}\ \bibnamefont
  {Miransky}}, \bibinfo {author} {\bibfnamefont {I.~A.}\ \bibnamefont
  {Shovkovy}},\ and\ \bibinfo {author} {\bibfnamefont {P.~O.}\ \bibnamefont
  {Sukhachov}},\ }\bibfield  {title} {\bibinfo {title} {Anomalous
  thermoelectric phenomena in lattice models of {multi-Weyl} semimetals},\
  }\href@noop {} {\bibfield  {journal} {\bibinfo  {journal} {Phys. Rev. B}\
  }\textbf {\bibinfo {volume} {96}},\ \bibinfo {pages} {155138} (\bibinfo
  {year} {2017})}\BibitemShut {NoStop}%
\bibitem [{\citenamefont {Gorbar}\ \emph {et~al.}(2018)\citenamefont {Gorbar},
  \citenamefont {Miransky}, \citenamefont {Shovkovy},\ and\ \citenamefont
  {Sukhachov}}]{Gorbar2018}%
  \BibitemOpen
  \bibfield  {author} {\bibinfo {author} {\bibfnamefont {E.~V.}\ \bibnamefont
  {Gorbar}}, \bibinfo {author} {\bibfnamefont {V.~A.}\ \bibnamefont
  {Miransky}}, \bibinfo {author} {\bibfnamefont {I.~A.}\ \bibnamefont
  {Shovkovy}},\ and\ \bibinfo {author} {\bibfnamefont {P.~O.}\ \bibnamefont
  {Sukhachov}},\ }\bibfield  {title} {\bibinfo {title} {Consistent hydrodynamic
  theory of chiral electrons in {Weyl} semimetals},\ }\href
  {https://journals.aps.org/prb/abstract/10.1103/PhysRevB.97.121105} {\bibfield
   {journal} {\bibinfo  {journal} {Phys. Rev. B}\ }\textbf {\bibinfo {volume}
  {97}},\ \bibinfo {pages} {121105(R)} (\bibinfo {year} {2018})}\BibitemShut
  {NoStop}%
\bibitem [{\citenamefont {Lucas}\ \emph {et~al.}(2016)\citenamefont {Lucas},
  \citenamefont {Davison},\ and\ \citenamefont {Sachdev}}]{Lucas2016}%
  \BibitemOpen
  \bibfield  {author} {\bibinfo {author} {\bibfnamefont {A.}~\bibnamefont
  {Lucas}}, \bibinfo {author} {\bibfnamefont {R.~A.}\ \bibnamefont {Davison}},\
  and\ \bibinfo {author} {\bibfnamefont {S.}~\bibnamefont {Sachdev}},\
  }\bibfield  {title} {\bibinfo {title} {Hydrodynamic theory of thermoelectric
  transport and negative magnetoresistance in {Weyl} semimetals},\ }\href
  {https://doi.org/10.1073/pnas.1608881113} {\bibfield  {journal} {\bibinfo
  {journal} {Proc. Natl. Acad. Sci. U.S.A.}\ }\textbf {\bibinfo {volume}
  {113}},\ \bibinfo {pages} {9463} (\bibinfo {year} {2016})}\BibitemShut
  {NoStop}%
\bibitem [{\citenamefont {Abrikosov}\ and\ \citenamefont
  {Beneslavski\u{i}}(1971)}]{Abrikosov1971}%
  \BibitemOpen
  \bibfield  {author} {\bibinfo {author} {\bibfnamefont {A.~A.}\ \bibnamefont
  {Abrikosov}}\ and\ \bibinfo {author} {\bibfnamefont {S.~D.}\ \bibnamefont
  {Beneslavski\u{i}}},\ }\bibfield  {title} {\bibinfo {title} {Possible
  existence of substances intermediate between metals and dielectrics},\
  }\href@noop {} {\bibfield  {journal} {\bibinfo  {journal} {Sov. Phys. JETP}\
  }\textbf {\bibinfo {volume} {32}},\ \bibinfo {pages} {699} (\bibinfo {year}
  {1971})}\BibitemShut {NoStop}%
\bibitem [{\citenamefont {Landau}\ and\ \citenamefont
  {Lifshitz}(2013{\natexlab{a}})}]{Landau2003}%
  \BibitemOpen
  \bibfield  {author} {\bibinfo {author} {\bibfnamefont {L.~D.}\ \bibnamefont
  {Landau}}\ and\ \bibinfo {author} {\bibfnamefont {E.~M.}\ \bibnamefont
  {Lifshitz}},\ }\href@noop {} {\emph {\bibinfo {title} {Fluid Mechanics:
  Volume 6}}},\ Vol.~\bibinfo {volume} {6}\ (\bibinfo  {publisher} {Elsevier},\
  \bibinfo {year} {2013})\BibitemShut {NoStop}%
\bibitem [{\citenamefont {Narozhny}\ \emph {et~al.}(2017)\citenamefont
  {Narozhny}, \citenamefont {Gornyi}, \citenamefont {Mirlin},\ and\
  \citenamefont {Schmalian}}]{Narozhny2017}%
  \BibitemOpen
  \bibfield  {author} {\bibinfo {author} {\bibfnamefont {B.~N.}\ \bibnamefont
  {Narozhny}}, \bibinfo {author} {\bibfnamefont {I.~V.}\ \bibnamefont
  {Gornyi}}, \bibinfo {author} {\bibfnamefont {A.~D.}\ \bibnamefont {Mirlin}},\
  and\ \bibinfo {author} {\bibfnamefont {J.}~\bibnamefont {Schmalian}},\
  }\bibfield  {title} {\bibinfo {title} {Hydrodynamic approach to electronic
  transport in graphene},\ }\href@noop {} {\bibfield  {journal} {\bibinfo
  {journal} {Ann. Phys.}\ }\textbf {\bibinfo {volume} {529}},\ \bibinfo {pages}
  {1700043} (\bibinfo {year} {2017})}\BibitemShut {NoStop}%
\bibitem [{\citenamefont {Narozhny}(2019)}]{Narozhny2019}%
  \BibitemOpen
  \bibfield  {author} {\bibinfo {author} {\bibfnamefont {B.~N.}\ \bibnamefont
  {Narozhny}},\ }\bibfield  {title} {\bibinfo {title} {Electronic hydrodynamics
  in graphene},\ }\href {https://doi.org/10.1016/J.AOP.2019.167979} {\bibfield
  {journal} {\bibinfo  {journal} {Ann. Phys.}\ }\textbf {\bibinfo {volume}
  {411}},\ \bibinfo {pages} {167979} (\bibinfo {year} {2019})}\BibitemShut
  {NoStop}%
\bibitem [{\citenamefont {Hartnoll}\ \emph {et~al.}(2007)\citenamefont
  {Hartnoll}, \citenamefont {Kovtun}, \citenamefont {M{\"u}ller},\ and\
  \citenamefont {Sachdev}}]{Hartnoll2007}%
  \BibitemOpen
  \bibfield  {author} {\bibinfo {author} {\bibfnamefont {S.~A.}\ \bibnamefont
  {Hartnoll}}, \bibinfo {author} {\bibfnamefont {P.~K.}\ \bibnamefont
  {Kovtun}}, \bibinfo {author} {\bibfnamefont {M.}~\bibnamefont {M{\"u}ller}},\
  and\ \bibinfo {author} {\bibfnamefont {S.}~\bibnamefont {Sachdev}},\
  }\bibfield  {title} {\bibinfo {title} {Theory of the {Nernst} effect near
  quantum phase transitions in condensed matter and in dyonic black holes},\
  }\href@noop {} {\bibfield  {journal} {\bibinfo  {journal} {Phys. Rev. B}\
  }\textbf {\bibinfo {volume} {76}},\ \bibinfo {pages} {144502} (\bibinfo
  {year} {2007})}\BibitemShut {NoStop}%
\bibitem [{\citenamefont {M{\"u}ller}\ \emph {et~al.}(2008)\citenamefont
  {M{\"u}ller}, \citenamefont {Fritz},\ and\ \citenamefont
  {Sachdev}}]{Muller2008}%
  \BibitemOpen
  \bibfield  {author} {\bibinfo {author} {\bibfnamefont {M.}~\bibnamefont
  {M{\"u}ller}}, \bibinfo {author} {\bibfnamefont {L.}~\bibnamefont {Fritz}},\
  and\ \bibinfo {author} {\bibfnamefont {S.}~\bibnamefont {Sachdev}},\
  }\bibfield  {title} {\bibinfo {title} {Quantum-critical relativistic
  magnetotransport in graphene},\ }\href@noop {} {\bibfield  {journal}
  {\bibinfo  {journal} {Phys. Rev. B}\ }\textbf {\bibinfo {volume} {78}},\
  \bibinfo {pages} {115406} (\bibinfo {year} {2008})}\BibitemShut {NoStop}%
\bibitem [{\citenamefont {M{\"u}ller}\ \emph {et~al.}(2009)\citenamefont
  {M{\"u}ller}, \citenamefont {Fritz}, \citenamefont {Sachdev},\ and\
  \citenamefont {Schmalian}}]{Miiller2009}%
  \BibitemOpen
  \bibfield  {author} {\bibinfo {author} {\bibfnamefont {M.}~\bibnamefont
  {M{\"u}ller}}, \bibinfo {author} {\bibfnamefont {L.}~\bibnamefont {Fritz}},
  \bibinfo {author} {\bibfnamefont {S.}~\bibnamefont {Sachdev}},\ and\ \bibinfo
  {author} {\bibfnamefont {J.}~\bibnamefont {Schmalian}},\ }\bibfield  {title}
  {\bibinfo {title} {Relativistic magnetotransport in graphene},\ }\href@noop
  {} {\bibfield  {journal} {\bibinfo  {journal} {AIP Conf. Proc}\ }\textbf
  {\bibinfo {volume} {1134}},\ \bibinfo {pages} {170} (\bibinfo {year}
  {2009})}\BibitemShut {NoStop}%
\bibitem [{\citenamefont {Foster}\ and\ \citenamefont
  {Aleiner}(2009)}]{Foster2009}%
  \BibitemOpen
  \bibfield  {author} {\bibinfo {author} {\bibfnamefont {M.~S.}\ \bibnamefont
  {Foster}}\ and\ \bibinfo {author} {\bibfnamefont {I.~L.}\ \bibnamefont
  {Aleiner}},\ }\bibfield  {title} {\bibinfo {title} {Slow imbalance relaxation
  and thermoelectric transport in graphene},\ }\href
  {https://doi.org/10.1103/PhysRevB.79.085415} {\bibfield  {journal} {\bibinfo
  {journal} {Phys. Rev. B}\ }\textbf {\bibinfo {volume} {79}},\ \bibinfo
  {pages} {085415} (\bibinfo {year} {2009})}\BibitemShut {NoStop}%
\bibitem [{\citenamefont {Son}\ and\ \citenamefont {Spivak}(2013)}]{Son2013}%
  \BibitemOpen
  \bibfield  {author} {\bibinfo {author} {\bibfnamefont {D.~T.}\ \bibnamefont
  {Son}}\ and\ \bibinfo {author} {\bibfnamefont {B.~Z.}\ \bibnamefont
  {Spivak}},\ }\bibfield  {title} {\bibinfo {title} {Chiral anomaly and
  classical negative magnetoresistance of {Weyl} metals},\ }\href@noop {}
  {\bibfield  {journal} {\bibinfo  {journal} {Phys. Rev. B}\ }\textbf {\bibinfo
  {volume} {88}},\ \bibinfo {pages} {104412} (\bibinfo {year}
  {2013})}\BibitemShut {NoStop}%
\bibitem [{\citenamefont {Cooper}\ \emph {et~al.}(1997)\citenamefont {Cooper},
  \citenamefont {Halperin},\ and\ \citenamefont {Ruzin}}]{Cooper1997}%
  \BibitemOpen
  \bibfield  {author} {\bibinfo {author} {\bibfnamefont {N.~R.}\ \bibnamefont
  {Cooper}}, \bibinfo {author} {\bibfnamefont {B.~I.}\ \bibnamefont
  {Halperin}},\ and\ \bibinfo {author} {\bibfnamefont {I.~M.}\ \bibnamefont
  {Ruzin}},\ }\bibfield  {title} {\bibinfo {title} {Thermoelectric response of
  an interacting two-dimensional electron gas in a quantizing magnetic field},\
  }\href@noop {} {\bibfield  {journal} {\bibinfo  {journal} {Phys. Rev. B}\
  }\textbf {\bibinfo {volume} {55}},\ \bibinfo {pages} {2344} (\bibinfo {year}
  {1997})}\BibitemShut {NoStop}%
\bibitem [{\citenamefont {{Das Sarma}}\ \emph {et~al.}(2015)\citenamefont {{Das
  Sarma}}, \citenamefont {Hwang},\ and\ \citenamefont {Min}}]{Sarma2015}%
  \BibitemOpen
  \bibfield  {author} {\bibinfo {author} {\bibfnamefont {S.}~\bibnamefont {{Das
  Sarma}}}, \bibinfo {author} {\bibfnamefont {E.~H.}\ \bibnamefont {Hwang}},\
  and\ \bibinfo {author} {\bibfnamefont {H.}~\bibnamefont {Min}},\ }\bibfield
  {title} {\bibinfo {title} {Carrier screening, transport, and relaxation in
  three-dimensional {Dirac} semimetals},\ }\href@noop {} {\bibfield  {journal}
  {\bibinfo  {journal} {Phys. Rev. B}\ }\textbf {\bibinfo {volume} {91}},\
  \bibinfo {pages} {035201} (\bibinfo {year} {2015})}\BibitemShut {NoStop}%
\bibitem [{Note1()}]{Note1}%
  \BibitemOpen
  \bibinfo {note} {The right most expression in Eq. (\ref {eq:sigma_xx_infty})
  has a factor two (the number of Weyl nodes) compared to the single node
  result, due to the assumption of no internode scattering by the disorder.
  Therefore, the inverse of $\protect \mathaccentV {bar}016{\tau }_{\protect
  \textrm {\protect \ensuremath {\parallel }}}^{\protect \textrm {el}}$ scales
  as the density of states of a single node, while $\partial n/\partial \mu $
  in the expression for $\sigma _{xx}$ gives the total density of
  states.}\BibitemShut {Stop}%
\bibitem [{\citenamefont {Sinitsyn}\ \emph {et~al.}(2007)\citenamefont
  {Sinitsyn}, \citenamefont {MacDonald}, \citenamefont {Jungwirth},
  \citenamefont {Dugaev},\ and\ \citenamefont {Sinova}}]{Sinitsyn2007}%
  \BibitemOpen
  \bibfield  {author} {\bibinfo {author} {\bibfnamefont {N.~A.}\ \bibnamefont
  {Sinitsyn}}, \bibinfo {author} {\bibfnamefont {A.~H.}\ \bibnamefont
  {MacDonald}}, \bibinfo {author} {\bibfnamefont {T.}~\bibnamefont
  {Jungwirth}}, \bibinfo {author} {\bibfnamefont {V.~K.}\ \bibnamefont
  {Dugaev}},\ and\ \bibinfo {author} {\bibfnamefont {J.}~\bibnamefont
  {Sinova}},\ }\bibfield  {title} {\bibinfo {title} {Anomalous {Hall} effect in
  a two-dimensional {Dirac} band: The link between the {Kubo-Streda} formula
  and the semiclassical {Boltzmann} equation approach},\ }\href
  {https://doi.org/10.1103/PhysRevB.75.045315} {\bibfield  {journal} {\bibinfo
  {journal} {Phys. Rev. B}\ }\textbf {\bibinfo {volume} {75}},\ \bibinfo
  {pages} {045315} (\bibinfo {year} {2007})}\BibitemShut {NoStop}%
\bibitem [{\citenamefont {Ado}\ \emph {et~al.}(2015)\citenamefont {Ado},
  \citenamefont {Dmitriev}, \citenamefont {Ostrovsky},\ and\ \citenamefont
  {Titov}}]{Ado2015}%
  \BibitemOpen
  \bibfield  {author} {\bibinfo {author} {\bibfnamefont {I.~A.}\ \bibnamefont
  {Ado}}, \bibinfo {author} {\bibfnamefont {I.~A.}\ \bibnamefont {Dmitriev}},
  \bibinfo {author} {\bibfnamefont {P.~M.}\ \bibnamefont {Ostrovsky}},\ and\
  \bibinfo {author} {\bibfnamefont {M.}~\bibnamefont {Titov}},\ }\bibfield
  {title} {\bibinfo {title} {Anomalous {Hall} effect with massive {Dirac}
  fermions},\ }\href@noop {} {\bibfield  {journal} {\bibinfo  {journal} {EPL}\
  }\textbf {\bibinfo {volume} {111}},\ \bibinfo {pages} {37004} (\bibinfo
  {year} {2015})}\BibitemShut {NoStop}%
\bibitem [{\citenamefont {Smrcka}\ and\ \citenamefont
  {Streda}(1977)}]{Smrcka1977}%
  \BibitemOpen
  \bibfield  {author} {\bibinfo {author} {\bibfnamefont {L.}~\bibnamefont
  {Smrcka}}\ and\ \bibinfo {author} {\bibfnamefont {P.}~\bibnamefont
  {Streda}},\ }\bibfield  {title} {\bibinfo {title} {Transport coefficients in
  strong magnetic fields},\ }\href@noop {} {\bibfield  {journal} {\bibinfo
  {journal} {J. Phys. C: Solid State Phys.}\ }\textbf {\bibinfo {volume}
  {10}},\ \bibinfo {pages} {2153} (\bibinfo {year} {1977})}\BibitemShut
  {NoStop}%
\bibitem [{\citenamefont {Burkov}(2014)}]{Burkov2014}%
  \BibitemOpen
  \bibfield  {author} {\bibinfo {author} {\bibfnamefont {A.~A.}\ \bibnamefont
  {Burkov}},\ }\bibfield  {title} {\bibinfo {title} {Anomalous {Hall} effect in
  {Weyl} metals},\ }\href {https://doi.org/10.1103/PhysRevLett.113.187202}
  {\bibfield  {journal} {\bibinfo  {journal} {Phys. Rev. Lett.}\ }\textbf
  {\bibinfo {volume} {113}},\ \bibinfo {pages} {187202} (\bibinfo {year}
  {2014})}\BibitemShut {NoStop}%
\bibitem [{\citenamefont {Pesin}(2018)}]{Pesin2018}%
  \BibitemOpen
  \bibfield  {author} {\bibinfo {author} {\bibfnamefont {D.~A.}\ \bibnamefont
  {Pesin}},\ }\bibfield  {title} {\bibinfo {title} {Two-particle collisional
  coordinate shifts and hydrodynamic anomalous {Hall} effect in systems without
  {Lorentz} invariance},\ }\href@noop {} {\bibfield  {journal} {\bibinfo
  {journal} {Phys. Rev. Lett.}\ }\textbf {\bibinfo {volume} {121}},\ \bibinfo
  {pages} {226601} (\bibinfo {year} {2018})}\BibitemShut {NoStop}%
\bibitem [{\citenamefont {Glazov}\ and\ \citenamefont
  {Golub}(2022)}]{Glazov2022}%
  \BibitemOpen
  \bibfield  {author} {\bibinfo {author} {\bibfnamefont {M.~M.}\ \bibnamefont
  {Glazov}}\ and\ \bibinfo {author} {\bibfnamefont {L.~E.}\ \bibnamefont
  {Golub}},\ }\bibfield  {title} {\bibinfo {title} {Spin and valley {Hall}
  effects induced by asymmetric interparticle scattering},\ }\href@noop {}
  {\bibfield  {journal} {\bibinfo  {journal} {Phys. Rev. B}\ }\textbf {\bibinfo
  {volume} {106}},\ \bibinfo {pages} {235305} (\bibinfo {year}
  {2022})}\BibitemShut {NoStop}%
\bibitem [{\citenamefont {Steiner}\ \emph {et~al.}(2017)\citenamefont
  {Steiner}, \citenamefont {Andreev},\ and\ \citenamefont
  {Pesin}}]{Steiner2017}%
  \BibitemOpen
  \bibfield  {author} {\bibinfo {author} {\bibfnamefont {J.~F.}\ \bibnamefont
  {Steiner}}, \bibinfo {author} {\bibfnamefont {A.~V.}\ \bibnamefont
  {Andreev}},\ and\ \bibinfo {author} {\bibfnamefont {D.~A.}\ \bibnamefont
  {Pesin}},\ }\bibfield  {title} {\bibinfo {title} {Anomalous {Hall} effect in
  {Type-I} {Weyl} metals},\ }\href
  {https://doi.org/10.1103/PhysRevLett.119.036601} {\bibfield  {journal}
  {\bibinfo  {journal} {Phys. Rev. Lett.}\ }\textbf {\bibinfo {volume} {119}},\
  \bibinfo {pages} {036601} (\bibinfo {year} {2017})}\BibitemShut {NoStop}%
\bibitem [{\citenamefont {Zhang}\ \emph {et~al.}(2023)\citenamefont {Zhang},
  \citenamefont {Wang},\ and\ \citenamefont {Chen}}]{zhang2023disorder}%
  \BibitemOpen
  \bibfield  {author} {\bibinfo {author} {\bibfnamefont {J.-X.}\ \bibnamefont
  {Zhang}}, \bibinfo {author} {\bibfnamefont {Z.-Y.}\ \bibnamefont {Wang}},\
  and\ \bibinfo {author} {\bibfnamefont {W.}~\bibnamefont {Chen}},\ }\bibfield
  {title} {\bibinfo {title} {Disorder-induced anomalous {Hall} effect in
  {Type-I Weyl} metals: Connection between the {Kubo-Streda} formula in the
  spin and chiral basis},\ }\href@noop {} {\bibfield  {journal} {\bibinfo
  {journal} {Phys. Rev. B}\ }\textbf {\bibinfo {volume} {107}},\ \bibinfo
  {pages} {125106} (\bibinfo {year} {2023})}\BibitemShut {NoStop}%
\bibitem [{\citenamefont {Ferreiros}\ \emph {et~al.}(2017)\citenamefont
  {Ferreiros}, \citenamefont {Zyuzin},\ and\ \citenamefont
  {Bardarson}}]{Ferreiros2017}%
  \BibitemOpen
  \bibfield  {author} {\bibinfo {author} {\bibfnamefont {Y.}~\bibnamefont
  {Ferreiros}}, \bibinfo {author} {\bibfnamefont {A.~A.}\ \bibnamefont
  {Zyuzin}},\ and\ \bibinfo {author} {\bibfnamefont {J.~H.}\ \bibnamefont
  {Bardarson}},\ }\bibfield  {title} {\bibinfo {title} {Anomalous {Nernst} and
  thermal {Hall} effects in tilted {Weyl} semimetals},\ }\href
  {https://doi.org/10.1103/PhysRevB.96.115202} {\bibfield  {journal} {\bibinfo
  {journal} {Phys. Rev. B}\ }\textbf {\bibinfo {volume} {96}},\ \bibinfo
  {pages} {115202} (\bibinfo {year} {2017})}\BibitemShut {NoStop}%
\bibitem [{\citenamefont {Saha}\ and\ \citenamefont {Tewari}(2018)}]{Saha2018}%
  \BibitemOpen
  \bibfield  {author} {\bibinfo {author} {\bibfnamefont {S.}~\bibnamefont
  {Saha}}\ and\ \bibinfo {author} {\bibfnamefont {S.}~\bibnamefont {Tewari}},\
  }\bibfield  {title} {\bibinfo {title} {Anomalous {Nernst} effect in type-{II}
  {Weyl} semimetals},\ }\href {https://doi.org/10.1140/epjb/e2017-80437-4}
  {\bibfield  {journal} {\bibinfo  {journal} {Eur. Phys. J. B}\ }\textbf
  {\bibinfo {volume} {91}},\ \bibinfo {pages} {1} (\bibinfo {year}
  {2018})}\BibitemShut {NoStop}%
\bibitem [{\citenamefont {Zyuzina}\ and\ \citenamefont
  {Tiwari}(2016)}]{Zyuzina2016}%
  \BibitemOpen
  \bibfield  {author} {\bibinfo {author} {\bibfnamefont {A.~A.}\ \bibnamefont
  {Zyuzina}}\ and\ \bibinfo {author} {\bibfnamefont {R.~P.}\ \bibnamefont
  {Tiwari}},\ }\bibfield  {title} {\bibinfo {title} {Intrinsic anomalous {Hall}
  effect in type-{II} {Weyl} semimetals},\ }\href
  {https://doi.org/10.1134/S002136401611014X} {\bibfield  {journal} {\bibinfo
  {journal} {JETP Lett.}\ }\textbf {\bibinfo {volume} {103}},\ \bibinfo {pages}
  {717} (\bibinfo {year} {2016})}\BibitemShut {NoStop}%
\bibitem [{\citenamefont {Li}\ \emph {et~al.}(2020)\citenamefont {Li},
  \citenamefont {Koo}, \citenamefont {Ning}, \citenamefont {Li}, \citenamefont
  {Miao}, \citenamefont {Min}, \citenamefont {Zhu}, \citenamefont {Wang},
  \citenamefont {Alem}, \citenamefont {Liu}, \citenamefont {Mao},\ and\
  \citenamefont {Yan}}]{Li2020}%
  \BibitemOpen
  \bibfield  {author} {\bibinfo {author} {\bibfnamefont {P.}~\bibnamefont
  {Li}}, \bibinfo {author} {\bibfnamefont {J.}~\bibnamefont {Koo}}, \bibinfo
  {author} {\bibfnamefont {W.}~\bibnamefont {Ning}}, \bibinfo {author}
  {\bibfnamefont {J.}~\bibnamefont {Li}}, \bibinfo {author} {\bibfnamefont
  {L.}~\bibnamefont {Miao}}, \bibinfo {author} {\bibfnamefont {L.}~\bibnamefont
  {Min}}, \bibinfo {author} {\bibfnamefont {Y.}~\bibnamefont {Zhu}}, \bibinfo
  {author} {\bibfnamefont {Y.}~\bibnamefont {Wang}}, \bibinfo {author}
  {\bibfnamefont {N.}~\bibnamefont {Alem}}, \bibinfo {author} {\bibfnamefont
  {C.~X.}\ \bibnamefont {Liu}}, \bibinfo {author} {\bibfnamefont
  {Z.}~\bibnamefont {Mao}},\ and\ \bibinfo {author} {\bibfnamefont
  {B.}~\bibnamefont {Yan}},\ }\bibfield  {title} {\bibinfo {title} {Giant room
  temperature anomalous {Hall} effect and tunable topology in a ferromagnetic
  topological semimetal {Co$_2$MnAl}},\ }\href@noop {} {\bibfield  {journal}
  {\bibinfo  {journal} {Nat. Commun.}\ }\textbf {\bibinfo {volume} {11}},\
  \bibinfo {pages} {1} (\bibinfo {year} {2020})}\BibitemShut {NoStop}%
\bibitem [{\citenamefont {Chen}\ \emph {et~al.}(2021)\citenamefont {Chen},
  \citenamefont {Li}, \citenamefont {Ding}, \citenamefont {Zhang},
  \citenamefont {Liu},\ and\ \citenamefont {Wang}}]{Chen2021}%
  \BibitemOpen
  \bibfield  {author} {\bibinfo {author} {\bibfnamefont {J.}~\bibnamefont
  {Chen}}, \bibinfo {author} {\bibfnamefont {H.}~\bibnamefont {Li}}, \bibinfo
  {author} {\bibfnamefont {B.}~\bibnamefont {Ding}}, \bibinfo {author}
  {\bibfnamefont {H.}~\bibnamefont {Zhang}}, \bibinfo {author} {\bibfnamefont
  {E.}~\bibnamefont {Liu}},\ and\ \bibinfo {author} {\bibfnamefont
  {W.}~\bibnamefont {Wang}},\ }\bibfield  {title} {\bibinfo {title} {Large
  anomalous {Hall} angle in a topological semimetal candidate {TbPtBi}},\
  }\href@noop {} {\bibfield  {journal} {\bibinfo  {journal} {Appl. Phys.
  Lett.}\ }\textbf {\bibinfo {volume} {118}},\ \bibinfo {pages} {031901}
  (\bibinfo {year} {2021})}\BibitemShut {NoStop}%
\bibitem [{\citenamefont {Behnia}\ \emph {et~al.}(2004)\citenamefont {Behnia},
  \citenamefont {Jaccard},\ and\ \citenamefont {Flouquet}}]{Behnia2004}%
  \BibitemOpen
  \bibfield  {author} {\bibinfo {author} {\bibfnamefont {K.}~\bibnamefont
  {Behnia}}, \bibinfo {author} {\bibfnamefont {D.}~\bibnamefont {Jaccard}},\
  and\ \bibinfo {author} {\bibfnamefont {J.}~\bibnamefont {Flouquet}},\
  }\bibfield  {title} {\bibinfo {title} {On the thermoelectricity of correlated
  electrons in the zero-temperature limit},\ }\href@noop {} {\bibfield
  {journal} {\bibinfo  {journal} {J. Phys. Condens. Matter}\ }\textbf {\bibinfo
  {volume} {16}},\ \bibinfo {pages} {5187} (\bibinfo {year}
  {2004})}\BibitemShut {NoStop}%
\bibitem [{\citenamefont {Floyd}(1997)}]{Floyd1997}%
  \BibitemOpen
  \bibfield  {author} {\bibinfo {author} {\bibfnamefont {T.~L.}\ \bibnamefont
  {Floyd}},\ }\href@noop {} {\emph {\bibinfo {title} {Principles of Electric
  Circuits}}},\ \bibinfo {edition} {5th}\ ed.\ (\bibinfo  {publisher} {Prentice
  Hall},\ \bibinfo {year} {1997})\BibitemShut {NoStop}%
\bibitem [{\citenamefont {Principi}\ and\ \citenamefont
  {Vignale}(2015)}]{Principi2015}%
  \BibitemOpen
  \bibfield  {author} {\bibinfo {author} {\bibfnamefont {A.}~\bibnamefont
  {Principi}}\ and\ \bibinfo {author} {\bibfnamefont {G.}~\bibnamefont
  {Vignale}},\ }\bibfield  {title} {\bibinfo {title} {Violation of the
  {Wiedemann-Franz} law in hydrodynamic electron liquids},\ }\href@noop {}
  {\bibfield  {journal} {\bibinfo  {journal} {Phys. Rev. Lett.}\ }\textbf
  {\bibinfo {volume} {115}},\ \bibinfo {pages} {056603} (\bibinfo {year}
  {2015})}\BibitemShut {NoStop}%
\bibitem [{\citenamefont {Tanaka}\ \emph {et~al.}(2020)\citenamefont {Tanaka},
  \citenamefont {Fujishiro}, \citenamefont {Mogi}, \citenamefont {Kaneko},
  \citenamefont {Yokosawa}, \citenamefont {Kanazawa}, \citenamefont {Minami},
  \citenamefont {Koretsune}, \citenamefont {Arita}, \citenamefont {Tarucha},
  \citenamefont {Yamamoto},\ and\ \citenamefont {Tokura}}]{Tanaka2020}%
  \BibitemOpen
  \bibfield  {author} {\bibinfo {author} {\bibfnamefont {M.}~\bibnamefont
  {Tanaka}}, \bibinfo {author} {\bibfnamefont {Y.}~\bibnamefont {Fujishiro}},
  \bibinfo {author} {\bibfnamefont {M.}~\bibnamefont {Mogi}}, \bibinfo {author}
  {\bibfnamefont {Y.}~\bibnamefont {Kaneko}}, \bibinfo {author} {\bibfnamefont
  {T.}~\bibnamefont {Yokosawa}}, \bibinfo {author} {\bibfnamefont
  {N.}~\bibnamefont {Kanazawa}}, \bibinfo {author} {\bibfnamefont
  {S.}~\bibnamefont {Minami}}, \bibinfo {author} {\bibfnamefont
  {T.}~\bibnamefont {Koretsune}}, \bibinfo {author} {\bibfnamefont
  {R.}~\bibnamefont {Arita}}, \bibinfo {author} {\bibfnamefont
  {S.}~\bibnamefont {Tarucha}}, \bibinfo {author} {\bibfnamefont
  {M.}~\bibnamefont {Yamamoto}},\ and\ \bibinfo {author} {\bibfnamefont
  {Y.}~\bibnamefont {Tokura}},\ }\bibfield  {title} {\bibinfo {title}
  {Topological {Kagome} magnet {Co$_3$Sn$_2$S$_2$} thin flakes with high
  electron mobility and large anomalous {Hall} effect},\ }\href@noop {}
  {\bibfield  {journal} {\bibinfo  {journal} {Nano Lett.}\ }\textbf {\bibinfo
  {volume} {20}},\ \bibinfo {pages} {7476} (\bibinfo {year}
  {2020})}\BibitemShut {NoStop}%
\bibitem [{Note2()}]{Note2}%
  \BibitemOpen
  \bibinfo {note} {Note that electron-electron internode scattering exists and
  is assumed to be much faster than the disorder scattering rate. Therefore,
  the assumption of no internode scattering by the disorder is not crucial, and
  relaxing it only modifies ${\tau }_{\parallel }^{\protect \textrm {el}}$ by a
  numerical factor}\BibitemShut {NoStop}%
\bibitem [{\citenamefont {Coulter}\ \emph {et~al.}(2018)\citenamefont
  {Coulter}, \citenamefont {Sundararaman},\ and\ \citenamefont
  {Narang}}]{Coulter2018}%
  \BibitemOpen
  \bibfield  {author} {\bibinfo {author} {\bibfnamefont {J.}~\bibnamefont
  {Coulter}}, \bibinfo {author} {\bibfnamefont {R.}~\bibnamefont
  {Sundararaman}},\ and\ \bibinfo {author} {\bibfnamefont {P.}~\bibnamefont
  {Narang}},\ }\bibfield  {title} {\bibinfo {title} {Microscopic origins of
  hydrodynamic transport in the type-{II} {Weyl} semimetal {WP$_2$}},\
  }\href@noop {} {\bibfield  {journal} {\bibinfo  {journal} {Phys. Rev. B}\
  }\textbf {\bibinfo {volume} {98}},\ \bibinfo {pages} {115130} (\bibinfo
  {year} {2018})}\BibitemShut {NoStop}%
\bibitem [{\citenamefont {Bernabeu}\ and\ \citenamefont
  {Cortijo}(2022)}]{Bernabeu2022}%
  \BibitemOpen
  \bibfield  {author} {\bibinfo {author} {\bibfnamefont {J.}~\bibnamefont
  {Bernabeu}}\ and\ \bibinfo {author} {\bibfnamefont {A.}~\bibnamefont
  {Cortijo}},\ }\bibfield  {title} {\bibinfo {title} {Phonon-mediated
  hydrodynamic transport in a {Weyl} semimetal},\ }\href@noop {} {\bibfield
  {journal} {\bibinfo  {journal} {arXiv preprint arXiv:2211.13245}\ } (\bibinfo
  {year} {2022})}\BibitemShut {NoStop}%
\bibitem [{\citenamefont {Osterhoudt}\ \emph {et~al.}(2021)\citenamefont
  {Osterhoudt}, \citenamefont {Wang}, \citenamefont {Garcia}, \citenamefont
  {Plisson}, \citenamefont {Gooth}, \citenamefont {Felser}, \citenamefont
  {Narang},\ and\ \citenamefont {Burch}}]{Osterhoudt2021}%
  \BibitemOpen
  \bibfield  {author} {\bibinfo {author} {\bibfnamefont {G.~B.}\ \bibnamefont
  {Osterhoudt}}, \bibinfo {author} {\bibfnamefont {Y.}~\bibnamefont {Wang}},
  \bibinfo {author} {\bibfnamefont {C.~A.~C.}\ \bibnamefont {Garcia}}, \bibinfo
  {author} {\bibfnamefont {V.~M.}\ \bibnamefont {Plisson}}, \bibinfo {author}
  {\bibfnamefont {J.}~\bibnamefont {Gooth}}, \bibinfo {author} {\bibfnamefont
  {C.}~\bibnamefont {Felser}}, \bibinfo {author} {\bibfnamefont
  {P.}~\bibnamefont {Narang}},\ and\ \bibinfo {author} {\bibfnamefont {K.~S.}\
  \bibnamefont {Burch}},\ }\bibfield  {title} {\bibinfo {title} {Evidence for
  dominant phonon-electron scattering in {Weyl} semimetal {WP$_2$}},\
  }\href@noop {} {\bibfield  {journal} {\bibinfo  {journal} {Phys. Rev. X}\
  }\textbf {\bibinfo {volume} {11}},\ \bibinfo {pages} {011017} (\bibinfo
  {year} {2021})}\BibitemShut {NoStop}%
\bibitem [{\citenamefont {Levchenko}\ and\ \citenamefont
  {Schmalian}(2020)}]{Levchenko2020}%
  \BibitemOpen
  \bibfield  {author} {\bibinfo {author} {\bibfnamefont {A.}~\bibnamefont
  {Levchenko}}\ and\ \bibinfo {author} {\bibfnamefont {J.}~\bibnamefont
  {Schmalian}},\ }\bibfield  {title} {\bibinfo {title} {Transport properties of
  strongly coupled electron-phonon liquids},\ }\href@noop {} {\bibfield
  {journal} {\bibinfo  {journal} {Ann. Phys.}\ }\textbf {\bibinfo {volume}
  {419}},\ \bibinfo {pages} {168218} (\bibinfo {year} {2020})}\BibitemShut
  {NoStop}%
\bibitem [{\citenamefont {Landau}\ and\ \citenamefont
  {Lifshitz}(2013{\natexlab{b}})}]{Landau1959}%
  \BibitemOpen
  \bibfield  {author} {\bibinfo {author} {\bibfnamefont {L.~D.}\ \bibnamefont
  {Landau}}\ and\ \bibinfo {author} {\bibfnamefont {E.~M.}\ \bibnamefont
  {Lifshitz}},\ }\href@noop {} {\emph {\bibinfo {title} {Statistical Physics:
  Volume 5}}}\ (\bibinfo  {publisher} {Elsevier},\ \bibinfo {year}
  {2013})\BibitemShut {NoStop}%
\bibitem [{\citenamefont {Lewin}(1981)}]{Lewin1981}%
  \BibitemOpen
  \bibfield  {author} {\bibinfo {author} {\bibfnamefont {L.}~\bibnamefont
  {Lewin}},\ }\href@noop {} {\emph {\bibinfo {title} {Polylogarithms and
  associated functions}}}\ (\bibinfo  {publisher} {Elsevier Science Limited},\
  \bibinfo {year} {1981})\BibitemShut {NoStop}%
\bibitem [{\citenamefont {Burkov}\ \emph {et~al.}(2011)\citenamefont {Burkov},
  \citenamefont {Hook},\ and\ \citenamefont {Balents}}]{Burkov2011}%
  \BibitemOpen
  \bibfield  {author} {\bibinfo {author} {\bibfnamefont {A.~A.}\ \bibnamefont
  {Burkov}}, \bibinfo {author} {\bibfnamefont {M.~D.}\ \bibnamefont {Hook}},\
  and\ \bibinfo {author} {\bibfnamefont {L.}~\bibnamefont {Balents}},\
  }\bibfield  {title} {\bibinfo {title} {Topological nodal semimetals},\ }\href
  {https://journals.aps.org/prb/abstract/10.1103/PhysRevB.84.235126} {\bibfield
   {journal} {\bibinfo  {journal} {Phys. Rev. B}\ }\textbf {\bibinfo {volume}
  {84}},\ \bibinfo {pages} {235126} (\bibinfo {year} {2011})}\BibitemShut
  {NoStop}%
\bibitem [{\citenamefont {Hosur}\ \emph {et~al.}(2012)\citenamefont {Hosur},
  \citenamefont {Parameswaran},\ and\ \citenamefont {Vishwanath}}]{Hosur2012}%
  \BibitemOpen
  \bibfield  {author} {\bibinfo {author} {\bibfnamefont {P.}~\bibnamefont
  {Hosur}}, \bibinfo {author} {\bibfnamefont {S.~A.}\ \bibnamefont
  {Parameswaran}},\ and\ \bibinfo {author} {\bibfnamefont {A.}~\bibnamefont
  {Vishwanath}},\ }\bibfield  {title} {\bibinfo {title} {Charge transport in
  {Weyl} semimetals},\ }\href
  {https://journals.aps.org/prl/abstract/10.1103/PhysRevLett.108.046602}
  {\bibfield  {journal} {\bibinfo  {journal} {Phys. Rev. Lett.}\ }\textbf
  {\bibinfo {volume} {108}},\ \bibinfo {pages} {046602} (\bibinfo {year}
  {2012})}\BibitemShut {NoStop}%
\bibitem [{\citenamefont {Strunk}(2021)}]{Strunk2021}%
  \BibitemOpen
  \bibfield  {author} {\bibinfo {author} {\bibfnamefont {C.}~\bibnamefont
  {Strunk}},\ }\bibfield  {title} {\bibinfo {title} {Quantum transport of
  particles and entropy},\ }\href@noop {} {\bibfield  {journal} {\bibinfo
  {journal} {Entropy}\ }\textbf {\bibinfo {volume} {23}},\ \bibinfo {pages}
  {1573} (\bibinfo {year} {2021})}\BibitemShut {NoStop}%
\bibitem [{\citenamefont {Lundgren}\ \emph {et~al.}(2014)\citenamefont
  {Lundgren}, \citenamefont {Laurell},\ and\ \citenamefont
  {Fiete}}]{Lundgren2014}%
  \BibitemOpen
  \bibfield  {author} {\bibinfo {author} {\bibfnamefont {R.}~\bibnamefont
  {Lundgren}}, \bibinfo {author} {\bibfnamefont {P.}~\bibnamefont {Laurell}},\
  and\ \bibinfo {author} {\bibfnamefont {G.~A.}\ \bibnamefont {Fiete}},\
  }\bibfield  {title} {\bibinfo {title} {Thermoelectric properties of {Weyl}
  and {Dirac} semimetals},\ }\href@noop {} {\bibfield  {journal} {\bibinfo
  {journal} {Phys. Rev. B}\ }\textbf {\bibinfo {volume} {90}},\ \bibinfo
  {pages} {165115} (\bibinfo {year} {2014})}\BibitemShut {NoStop}%
\end{thebibliography}%

\end{document}